\documentclass[a4paper,UKenglish,cleveref]{lipics-v2021}

\newif\ifarxivsubmit%
\arxivsubmittrue%
\newcommand{\arxivsubmittext}[2]{\ifarxivsubmit#1\else#2\fi}

\usepackage{mathtools}
\usepackage{esvect}
\renewcommand{\vec}[1]{\vv{#1}}

\usepackage{forest}
\tikzset{textselect/.style={font=\footnotesize\bfseries\boldmath, text=red}}
\tikzset{containerselect/.style={very thick, draw=red}}
\tikzset{select/.style={textselect, containerselect}}
\tikzset{node/.style={font=\footnotesize, draw=black, inner sep=1mm}}
\tikzset{container draw/.style={densely dotted}}
\tikzset{container/.style={node, container draw}}
\forestset{%
shortedges/.style={%
  for tree={l sep-=0.25cm,l-=0.25cm},
},
node/.style={%
  font=\footnotesize,
  draw=black,
  inner sep=1mm
},
goal/.style={%
  node
},
rule/.style={%
  node,
  dashed
},
goalcluster/.style={%
  node,
  dotted
},
proof/.style={%
  select,
  color=blue
},
child container/.style={tikz={\node() [container, fit to=children] {};}},
}

\usepackage{xspace}
\xspaceremoveexception{-}% repeated, for the sake of copy-pasting
\makeatletter
\renewcommand*\@xspace@hook{%
  \ifx\@let@token-%
    \expandafter\@xspace@dash@i
  \fi
}
\def\@xspace@dash@i-{\futurelet\@let@token\@xspace@dash@ii}
\def\@xspace@dash@ii{%
  \ifx\@let@token-%
  \else
    \unskip
  \fi
  -%
}
\makeatother

\definecolor{mlpurple}{rgb}{0.62, 0.12, 0.94}
\definecolor{mlgreen}{rgb}{0.00, 0.50, 0.00}
\definecolor{mlcomment}{rgb}{0.80, 0.00, 0.00}
\definecolor{mlantiquot}{rgb}{0.00, 0.0, 1.0}
\newcommand{\antiquotstyle}{\color{mlantiquot}}
\lstdefinelanguage{myml}{%
  language = ml,
  upquote=true,
  keywordstyle=\color{mlpurple},
  deletekeywords={struct,sig,end,let,in,end},
  keywords=[2]{struct,sig,end,let,in,end},
  keywordstyle=[2]\color{mlgreen},
  commentstyle=\color{mlcomment},
  moredelim=[is][\antiquotstyle]{[*}{*]},
  escapechar=`
}

\newcommand{\Zippy}{Zippy\xspace}
\newcommand{\isabelleml}{Isabelle\slash ML\xspace}
\newcommand{\isabellehol}{Isabelle\slash HOL\xspace}
\newcommand{\methodfont}[1]{\texttt{#1}}

\DeclarePairedDelimiter\parenths{(}{)}
\DeclarePairedDelimiter\brackets{[}{]}

\newcommand{\sauto}{\methodfont{sauto}\xspace}
\newcommand{\eauto}{\methodfont{eauto}\xspace}
\newcommand{\auto}{\methodfont{auto}\xspace}
\newcommand{\autotwo}{\methodfont{auto2}\xspace}
\newcommand{\aesop}{\methodfont{aesop}\xspace}
\newcommand{\grind}{\methodfont{grind}\xspace}
\newcommand{\psl}{\methodfont{PSL}\xspace}
\newcommand{\waterfall}{\methodfont{waterfall}\xspace}
\newcommand{\transfer}{\methodfont{transfer}\xspace}
\newcommand{\transport}{\methodfont{transport}\xspace}
\newcommand{\isaautoref}{\methodfont{autoref}\xspace}

\newenvironment{propenumerate}{\begin{enumerate}[(a)]}{\end{enumerate}}
\newcommand{\proptyref}[1]{Property~\eqref{#1}}

\newcommand{\mkconst}{\constfont{mk}}
\newcommand{\mkprefix}[1]{\mkconst#1}
\newcommand{\app}{\,}
\newcommand{\underscorematch}{\_}
\newcommand{\hasty}{:}
\newcommand{\holhasty}{:}
\newcommand{\modplus}{\oplus}
\newcommand{\modsub}{\ominus}
\newcommand{\opaccess}[2]{#1.#2}
\newcommand{\mvar}[1]{\mathord{?}#1}

\newcommand{\antiquot}[1]{{\color{mlantiquot}<#1>}}
\newcommand{\cartoucheleft}{\guilsinglleft}
\newcommand{\cartoucheight}{\guilsinglright}
\newcommand{\cartouches}[1]{\cartoucheleft #1\cartoucheight}
\newcommand{\eval}[1]{{\color{mlantiquot}\antiquot{eval}\cartouches{#1}}}
\newcommand{\pargs}[1]{{\color{mlantiquot}\antiquot{pargs}\cartouches{#1}}}
\newcommand{\npargs}{\antiquot{\#pargs}}
\newcommand{\inlineml}[1]{\lstinline[language=myml]{#1}}
\newcommand{\imap}[1]{{\color{mlantiquot}\antiquot{imap}\cartouches{#1}}}
\newcommand{\zargs}{\antiquot{zargs}}
\newcommand{\znargs}{\antiquot{\#zargs}}

\newcommand{\typefont}[1]{\mathsf{#1}}
\newcommand{\purefun}{\mathrel\typefont{\Rightarrow}}
\newcommand{\alphaty}{\typefont{\alpha}}
\newcommand{\betaty}{\typefont{\beta}}
\newcommand{\gammaty}{\typefont{\gamma}}
\newcommand{\listty}[1]{#1\app\typefont{list}}
\newcommand{\treety}[1]{#1\app\typefont{tree}}
\newcommand{\optionty}[1]{#1\app\typefont{option}}
\newcommand{\natty}{\mathbb{N}}
\newcommand{\intty}{\mathbb{Z}}
\newcommand{\stringty}{\typefont{string}}

\newcommand{\constfont}[1]{\mathsf{#1}}
\newcommand{\id}{\constfont{id}}
\newcommand{\cons}{\mathbin{\constfont{::}}}

\newcommand{\monadvar}{\typefont{m}}

\newcommand{\pure}{\constfont{pure}}
\newcommand{\bind}{\mathbin{\constfont{>\!\!\!>\mkern-6.7mu=}}}

\newcommand{\cat}{\rightarrow}
\newcommand{\compr}{\mathbin{\constfont{>\!\!\!>\!\!\!>}}}
\newcommand{\arr}{\constfont{arr}}

\newcommand{\arrsplit}{\mathbin{\constfont{*\!\!*\!\!*}}}
\newcommand{\hmovesym}{\circlearrowleft}
\newcommand{\hmove}[1]{#1^{\circlearrowleft}}
\newcommand{\hmoveexplicit}[2]{\hmove{#1}_{#2}}
\newcommand{\kleisli}[1]{\cat_{#1}}

\newcommand{\container}{\typefont{CO}}
\newcommand{\containern}[1]{\container_{#1}}

\newcommand{\node}{\typefont{N}}
\newcommand{\noden}[1]{\node_{#1}}
\newcommand{\ncontent}{\typefont{NC}}
\newcommand{\ncontentn}[1]{\ncontent_{#1}}
\newcommand{\nnext}{\typefont{NN}}
\newcommand{\nnextn}[1]{\nnext_{#1}}

\newcommand{\zipper}{\typefont{Z}}
\newcommand{\zippern}[1]{\zipper_{#1}}
\newcommand{\zcontent}{\typefont{ZC}}
\newcommand{\zcontentn}[1]{\zcontent_{#1}}
\newcommand{\zctxt}{\typefont{ZX}}
\newcommand{\zctxtn}[1]{\zctxt_{#1}}
\newcommand{\zup}{\constfont{up}}
\newcommand{\zdown}{\constfont{down}}
\newcommand{\zleft}{\constfont{left}}
\newcommand{\zright}{\constfont{right}}
\newcommand{\zzip}{\constfont{zip}}
\newcommand{\zunzip}{\constfont{unzip}}
\newcommand{\zdownn}[1]{\zdown_{#1}}
\newcommand{\zupn}[1]{\zup_{#1}}
\newcommand{\zipopn}[2]{\opaccess{\zippern{#1}}{#2}}

\newcommand{\anzcontainer}{\container'}
\newcommand{\anzcontainern}[1]{\anzcontainer_{#1}}
\newcommand{\anzctxt}{\zctxt'}
\newcommand{\anzctxtn}[1]{\anzctxt_{#1}}
\newcommand{\anzpzipper}{\typefont{P}\zipper}
\newcommand{\anzpzippern}[1]{\anzpzipper_{#1}}
\newcommand{\anzzipper}{\zipper'}
\newcommand{\anzzippern}[1]{\anzzipper_{#1}}
\newcommand{\zipperfrombasezipper}{\mkprefix{\anzzipper}}
\newcommand{\zipperfrombasezippern}[1]{\zipperfrombasezipper_{#1}}

\newcommand{\basezipperfromzippern}[1]{\zippern{#1}\constfont{from}\anzzippern{#1}}
\newcommand{\liftmove}{\constfont{liftMove}}
\newcommand{\liftmoven}[1]{\liftmove_{#1}}
\newcommand{\anzdownn}[1]{\zdownn{#1}}
\newcommand{\anzupn}[1]{\zupn{#1}}

\newcommand{\lens}[2]{(#1, #2)\app\typefont{lens}}
\newcommand{\getter}[2]{(#1, #2)\app\typefont{getter}}
\newcommand{\modifier}[2]{(#1, #2)\app\typefont{modifier}}

\newcommand{\state}[2]{(#1, #2)\app\typefont{state}}
\newcommand{\statetyvar}{\typefont{\sigma}}

\newcommand{\action}{\typefont{action}}
\newcommand{\tacticty}{\constfont{tactic}}
\newcommand{\tacaction}{\constfont{tacAction}}

\newcommand{\arrcatch}{\constfont{catch}}
\newcommand{\arrtry}{\constfont{try}}

\newcommand{\arrrepeat}{\constfont{repeat}}

\newcommand{\enumfirst}{\constfont{first}}
\newcommand{\enumnext}{\constfont{next}}
\newcommand{\maxaction}{\constfont{maxAction}}
\newcommand{\bestfirst}{\constfont{bestFirst}}
\newcommand{\totop}{\constfont{top}}
\newcommand{\applyaction}{\constfont{applyAction}}

\arxivsubmittext{%
\pdfoutput=1 %uncomment to ensure pdflatex processing (mandatatory e.g. to submit to arXiv)
\hideLIPIcs  %uncomment to remove references to LIPIcs series (logo, DOI, ...), e.g. when preparing a pre-final version to be uploaded to arXiv or another public repository
\nolinenumbers %uncomment to disable line numbering
}{}

\title{\Zippy -- Generic White-Box Proof Search with Zippers}

\author{Kevin Kappelmann}{Technical University of Munich, Boltzmannstrasse 3, 85748 Garching, Germany}{kevin.kappelmann@tum.de}{https://orcid.org/0000-0003-1421-6497}{}

\authorrunning{K. Kappelmann} %mandatory. First: Use abbreviated first/middle names. Second (only in severe cases): Use first author plus 'et al.'

\Copyright{Kevin Kappelmann} %mandatory, please use full first names. LIPIcs license is "CC-BY";  http://creativecommons.org/licenses/by/3.0/

\ccsdesc[500]{Theory of computation~Functional constructs}
\ccsdesc[300]{Theory of computation~Automated reasoning}
\ccsdesc[300]{Theory of computation~Logic and verification}

\keywords{Proof search, Functional programming, Arrows, Monads, Lenses, Software design} %mandatory; please add comma-separated list of keywords

\category{} %optional, e.g. invited paper

\relatedversion{} %TODO: optional, e.g. full version hosted on arXiv, HAL, or other respository/website
\arxivsubmittext{\supplement{\url{https://github.com/kappelmann/zippy-isabelle}}}{}
\EventEditors{John Q. Open and Joan R. Access}
\EventNoEds{2}
\EventLongTitle{42nd Conference on Very Important Topics (CVIT 2016)}
\EventShortTitle{CVIT 2016}
\EventAcronym{CVIT}
\EventYear{2016}
\EventDate{December 24--27, 2016}
\EventLocation{Little Whinging, United Kingdom}
\EventLogo{}
\SeriesVolume{42}
\ArticleNo{23}
\begin{document}

\maketitle

%mandatory: add short abstract of the document
\begin{abstract}
We present a framework for tree-based proof search, called \Zippy.
Unlike existing proof search tools,
\Zippy is largely independent of concrete search tree representations,
search-algorithms,
states and effects.
It is designed to create analysable and navigable proof searches
that are open to customisation and extensions by users.
\Zippy is founded on concepts from functional programming theory,
particularly zippers, arrows, monads, and lenses.
We implemented the framework in Isabelle's metaprogramming language \isabelleml.

\end{abstract}

\section{Introduction}\label{sec:intro}

The usability of proof assistants crucially depends on their proof automation.
There are different kinds of proof automation.
Some automation is \emph{domain-specific}, meaning that it operates on goals of a fixed form.
Examples include decision procedures for various theories, e.g.\ Presburger arithmetic.
Other automation is \emph{general-purpose}, meaning that it is (mostly) unconstrained in the form of its input goals.
Examples include
ACL2's~\cite{waterfall} \waterfall,
Coq's~\cite{coq} \eauto and \sauto~\cite{sauto},
Isabelle's~\cite{lcftoholauto} \auto,
Lean's \aesop~\cite{aesop}, and
PVS's \grind~\cite{grind}.

General-purpose automation commonly performs \emph{proof search}.
Proof search is the act of automatically exploring the space of valid proof derivations in one way or another.
Proof search tools typically try to find one or all proofs of a given goal.
Some proof search tools are \emph{black-box}, meaning that their exploration, i.e.\ the act of finding the proofs,
cannot readily be predicted or examined by users and programs.
Examples include tools integrating external solvers, so-called \emph{hammers}~\cite{hammers},
and machine-learning-based provers~\cite{htps}.
\emph{White-box} proof search tools, in contrast, provide various means to predict and analyse their proof exploration.
In this view, Isabelle's dominant general-purpose automation \auto is not truly white-box:
while it performs a fairly predictable depth-first search,
the search is implemented \emph{implicitly},
meaning that it uses recursion in Isabelle's metaprogramming language to perform the search and backtracking.
Coq's \eauto and \sauto similarly perform an implicit search.
Lean's \aesop, in contrast, implements its search \emph{explicitly},
meaning that it uses an explicit datatype (essentially, an AND/OR search tree)
modelling its exploration of proof derivations.
This search tree can be analysed as data in Lean itself.

Having access to proof explorations as data comes with many merits.
First, proof assistant users can analyse them.
For example, failing searches can be analysed to find gaps in the automation
or issues in the search algorithm -- such as looping cycles or unexpected search paths.
Successful searches, on the other hand, can be analysed to gain insight into a theorem's proof derivations and to understand what is happening under the hood of the automation.
Members of large-scale verification projects
noted the usefulness of such data for debugging, maintenance,
and onboarding of new proof engineers~\cite{manageproofs}.
Second, other tools can use the search data.
For example, machine-learning-based provers may be trained on it to find better search strategies,
select promising, though incomplete, paths in failed searches,
or suggest missing lemmas to make the search successful.
UI tools may use the data to visualise the search process interactively.
This way, search-based automation becomes more accessible to proof assistant users.

Another design dimension for general-purpose automation is
customisability and extensibility.
For instance, users of Isabelle's \auto, Coq's \sauto, and Lean's \aesop
can add and remove theorems that may be used during proof search,
register additional solvers for goals,
and limit the search depth.
Other options, however, are hard-wired,
such as the search algorithm itself in the case of \auto, \sauto,
ACL2' \waterfall, and PVS's \grind
or the datatypes and effects used during proof search.
Moreover, all mentioned tools are non-extensible
in the sense that there is no (simple) way to
add information to the proof search -- such
as additional caches, logs, and statistics --
without changing the respective tool's source code,
which we believe is ill-advised in practice:
First, most proof assistant users heavily rely on these tools and expect them to be stable.
They thus cannot be changed for user experiments.
Second, some modifications are non-compatible, such as competing representations of data stored during search.
Third, catering for many extensions in a single tool
may lead to a monolithic code base that tends to be hard to maintain and understand.
For example, Coq's \sauto already contains 35 hard-coded options and 14 action types,
ACL2 dozens of types of so-called hints and options for its \waterfall,
and Lean's \aesop 19 data slots for goals and 12 data slots for search rules at the time of writing.

We believe that too little attention has been paid to extensibility in the past.
One key reason why state-of-the-art tools are non-extensible
is that they are not based on sufficiently abstract concepts.
Instead, they are implemented against specifications and interfaces with little room for variation,
such as Lean's \aesop and Coq's \sauto,
or against no specification or interface at all,
such as Isabelle's \auto, ACL2's \waterfall, and PVS's \grind.
As a result, they cannot be easily extended
nor can their code be re-used by other proof search tools.

\paragraph*{Contributions and Outline}
To improve on the state-of-the-art,
we devise a customisable and extensible framework
for white-box, tree-based searches, called \Zippy,
and implement it in Isabelle's metaprogramming language.
Our contributions are as follows:
\begin{itemize}
\item We describe \Zippy's design in \cref{sec:zippy-design}.
\Zippy is a generic framework in that it is largely independent of concrete search tree representations, search strategies, and logics it is applied to.
At the same time, it does the bulk of heavy lifting needed for white-box, tree-based searches
while allowing customisations and extensions by users.
It does so by providing a specification for search tree navigation,
a mechanism to generate instances satisfying the specification,
and by using concepts and abstractions known from functional programming theory,
particularly zippers~\cite{zipper}, arrows~\cite{arrows}, monads~\cite{monads}, and lenses~\cite{lensorig,lenses}.
\item In \cref{sec:zippy-impl}, we implement
said concepts in \isabelleml~\cite{isabelleml}.
These concepts are prominent in languages with typeclasses, such as Haskell,
but typically not employed in ML-languages.
Though it is known that monads and the like can be modelled with ML's module system,
serious applications are largely missing.
We highlight a pitfall of this approach and how it can be taken care of in \isabelleml.
We also sketch how \Zippy's central concept of alternating zippers
can be defined in \isabelleml using its antiquotation mechanism~\cite{isabelleml}.
\item In \cref{sec:exmpl-applications},
we demonstrate one way of building a generic proof search tool with \Zippy in Isabelle.
We provide liftings to embed Isabelle tactics as search steps
and show how to implement
a best-first proof search in the style of Lean's \aesop.
% The example showcases some of the benefits of the framework's flexibility.
\end{itemize}
This article’s supplementary material
includes the implementation of \Zippy in \makebox{\isabelleml}
and an extended version of its example instantiation from \cref{sec:exmpl-applications}.
\cref{sec:related-work} contains related and future work.

\section{The Design of Zippy}\label{sec:zippy-design}

In this section, we devise the design of \Zippy.
We will provide an abstract specification for search tree navigation
and a mechanism to generate instances satisfying the specification
in \cref{sec:zippy-navigate}.
In \cref{sec:zippy-extensibility},
we will add concepts for extensible data manipulation.
\Zippy, then, is the sum of the devised specification, its instance generation,
and concepts for data manipulation.
Based on the considerations from \cref{sec:intro},
we first summarise the properties we desire for the framework:
\begin{propenumerate}
\item\label{propty:gen-purp} \Zippy should enable \emph{general-purpose automation}, i.e.\ automation that is unconstrained in the form of its input goals.
\item\label{propty:tree-based} \Zippy should support \emph{tree-based proof searches}, meaning that it should support proof explorations on search trees.
The reason for choosing tree-based searches is that proof trees are the de facto standard format for formal proof derivations.
Proof assistant users are thus familiar with them and
have an intuition of how various expansion strategies, possibly written by themselves,
affect searches on these trees.
This has also been noted by the developers of Lean's \aesop~\cite{aesop}.
\item\label{propty:white-box} \Zippy should enable \emph{white-box} searches, allowing users and other tools to analyse and modify proof explorations as data.
\item\label{propty:custom} \Zippy should be \emph{customisable}, i.e.\ it should be independent of concrete search tree representations,
search algorithms, states and effects.
\item\label{propty:extensible}\Zippy should be \emph{extensible}, allowing users to augment the search tree’s data
without modifying \Zippy{}’s source code
while being able to re-use existing code written for the simpler tree.
\end{propenumerate}
To achieve these properties, we consider Lean's \aesop
as a source of design inspiration.
It is based on an AND/OR tree,
where AND nodes contain the goals to be proven
and OR nodes contain possible steps to expand their parent goal nodes.
Next to this basic structure, its nodes contain various data,
e.g.\ user-provided success probabilities for OR nodes
and status flags (\texttt{proved}, \texttt{stuck}, \texttt{unknown}).
It performs a best-first search based on its OR nodes' success probabilities.
\cref{fig:aesop-tree-basic} shows a simplified example of this kind of search.

{%
\newcommand{\assumption}{Assm}
\newcommand{\contradiction}{Contradiction}
\newcommand{\goal}{$A\vdash (B\rightarrow C)\lor (A\land A)$}
\newcommand{\rulel}{${\lor}L$, 80\%}
\newcommand{\rulem}{${\lor}R$, 80\%}
\newcommand{\goall}{$A\vdash B\rightarrow C$}
\newcommand{\rulell}{${\rightarrow}I$, 60\%}
\newcommand{\goalm}{$A\vdash A\land A$}
\newcommand{\rulemm}{${\land}I$, 50\%}
\newcommand{\goalll}{$A,B\vdash C$}
\newcommand{\goalml}{$A \vdash A$}
\newcommand{\goalmr}{$A \vdash A$}
\newcommand{\rulemml}{\assumption,  30\%}
\newcommand{\rulemmr}{\assumption, 30\%}
\newcommand{\goalmlm}{$\top$}
\newcommand{\goalmrm}{$\top$}
\begin{figure}[t]
\begin{subfigure}[T]{0.33\textwidth}
\centering
\begin{forest}
shortedges [{\goal},  goal
  [{\rulel}, rule, select]
  [{\rulem}, rule]
]
\end{forest}
\end{subfigure}\hfill
\begin{subfigure}[T]{0.33\textwidth}
\centering
\begin{forest}
shortedges [{\goal},  goal
   [{\rulel}, rule [
     {\goall}, goal
       [{\rulell}, rule]]
   ]
   [{\rulem}, rule, select]
 ]
\end{forest}
\end{subfigure}\hfill
\begin{subfigure}[T]{0.33\textwidth}
\centering
\begin{forest}
shortedges [{\goal},  goal
   [{\rulel}, rule [
     {\goall}, goal
       [{\rulell}, rule, select]]
   ]
   [{\rulem}, rule [
     {\goalm}, goal
       [{\rulemm}, rule]]
   ]
 ]
\end{forest}
\end{subfigure}

\vspace{\baselineskip}
\begin{subfigure}[T]{0.5\textwidth}
\centering
\begin{forest}
shortedges [{\goal},  goal
   [{\rulel}, rule [
     {\goall}, goal
       [{\rulell}, rule
         [{\goalll}, goal
         ]
       ]]
   ]
   [{\rulem}, rule [
     {\goalm}, goal
       [{\rulemm}, rule, select]]
   ]
 ]
\end{forest}
\end{subfigure}\hfill
\begin{subfigure}[T]{0.5\textwidth}
\centering
\begin{forest}
shortedges [{\goal},  goal
   [{\rulel}, rule [
     {\goall}, goal
       [{\rulell}, rule
         [{\goalll}, goal
         ]
       ]]
   ]
   [{\rulem}, rule [
     {\goalm}, goal
       [{\rulemm}, rule
         [{\goalml}, goal
           [{\rulemml}, rule, select]
         ]
         [{\goalmr}, goal
           [{\rulemmr}, rule]
         ]
       ]]
   ]
]
\end{forest}
\end{subfigure}

\vspace{\baselineskip}
\begin{subfigure}[T]{0.5\textwidth}
\centering
\begin{forest}
shortedges [{\goal},  goal
   [{\rulel}, rule [
     {\goall}, goal
       [{\rulell}, rule
         [{\goalll}, goal
         ]
       ]]
   ]
   [{\rulem}, rule [
     {\goalm}, goal
       [{\rulemm}, rule
         [{\goalml}, goal
           [{\rulemml}, rule
             [{\goalmlm}, goal]
           ]
         ]
         [{\goalmr}, goal
           [{\rulemmr}, rule, select]
         ]
       ]]
   ]
]
\end{forest}
\end{subfigure}\hfill
\begin{subfigure}[T]{0.5\textwidth}
\centering
\begin{forest}
shortedges [{\goal},  goal, proof
   [{\rulel}, rule [
     {\goall}, goal
       [{\rulell}, rule
         [{\goalll}, goal
         ]
       ]]
   ]
   [{\rulem}, rule, proof [
     {\goalm}, goal, proof
       [{\rulemm}, rule, proof
         [{\goalml}, goal, proof
           [{\rulemml}, rule, proof
             [{\goalmlm}, goal, proof]
           ]
         ]
         [{\goalmr}, goal, proof
           [{\rulemmr}, rule, proof
             [{\goalmrm}, goal, proof]
           ]
         ]
       ]]
   ]
]
\end{forest}
\end{subfigure}
\caption{A step-by-step best-first proof search, from top-left to bottom-right. Goal nodes are solid and rule nodes are dashed, containing a rule ($\lor L$, \assumption, $\land I$, etc.) to expand their parent and a success probability. Selected rules nodes are bold and red. The resulting proof is bold and blue.}\label{fig:aesop-tree-basic}
\end{figure}
The example is quite paradigmatic for tree-based searches.
Here are some observations:
\begin{itemize}
\item The search tree consists of several node types $\vec{\alphaty}\app\noden{i}$,
where $\vec{\alphaty}=(\alphaty_1,\dotsc,\alphaty_n)$ is a (possibly empty) vector of polymorphic type arguments.
In the case of \cref{fig:aesop-tree-basic}, there are two node types: goals and rules.
\item Each node stores some content. Nodes of the same type $\vec{\alphaty}\app\noden{i}$ store content of the same type $\vec{\alphaty}\app\ncontentn{i}$.
Nodes of different types may store content of different types.
\item Each node of one type points to (possibly zero) children of another node type.
These children may be described by a type $\vec{\alphaty}\app\nnextn{i}$.
\item The node types are \emph{alternating},
meaning that the order of node types is fixed and cyclic.
In the example, goal nodes are followed by rule nodes, and vice versa.
\end{itemize}
We now define node types formally:
\begin{definition}[Nodes]
A \emph{node type} $\vec{\alphaty}\app\node=\vec{\alphaty}\app\ncontent\times\vec{\alphaty}\app\nnext$
consists of a \emph{(node) content type} $\vec{\alphaty}\app\ncontent$
and a \emph{(node) next type} $\vec{\alphaty}\app\nnext$.
\end{definition}
Based on above observations,
we propose an abstract search tree model as described in \cref{fig:gen-search-tree}.
The resulting model is modular and independent of concrete datatype representations.
It is parametrised by a number $n$,
container types $\vec{\alphaty}\app\containern{i}$, and content types $\vec{\alphaty}\app\ncontentn{i}$ for $1\leq i\leq n$.
\begin{figure}[t]
\begin{subfigure}[T]{0.20\textwidth}
\centering
\renewcommand{\goal}{$A\vdash C\lor (A\land A)$}
\renewcommand{\goall}{$A\vdash C$}
\begin{forest}
shortedges [{\goal}, name=root, goal, textselect,
   [{\rulel}, rule, textselect [
     {\goall}, goal, textselect]
   ]
   [{\rulem}, rule, textselect [
     {\goalm}, goal, textselect]
   ]
]
\end{forest}
\end{subfigure}\hfill
\renewcommand{\goal}{$\ncontentn{1}\mid \nnextn{1}$}
\renewcommand{\goall}{\goal}
\renewcommand{\goalm}{\goal}
\renewcommand{\rulel}{$\ncontentn{2}\mid \nnextn{2}$}
\renewcommand{\rulem}{\rulel}
\begin{subfigure}[T]{0.26\textwidth}
\centering
\begin{forest}
for tree={no edge},
shortedges [{\goal}, name=root, goal
   [{\rulel}, name=rulel, node [
     {\goall}, name=goall, node]
   ]
   [{\rulem}, name=ruler, node [
     {\goalm}, name=goalm, node]
   ]
]
\draw [select]([xshift=-3mm]root.south east)--(rulel.north);
\draw [select]([xshift=-3mm]root.south east)--(ruler.north);
\draw [select]([xshift=-3mm]rulel.south east)--(goall.north);
\draw [select]([xshift=-3mm]ruler.south east)--(goalm.north);
\end{forest}
\end{subfigure}\hfill
\begin{subfigure}[T]{0.28\textwidth}
\centering
\newbox\nodeboxone
\setbox\nodeboxone=\hbox{%
\begin{forest}
for tree={no edge},
[{\goal}, grow=east, node]
\end{forest}}
\newbox\nodeboxtwo
\setbox\nodeboxtwo=\hbox{%
\begin{forest}
for tree={no edge},
[{\rulel}, grow=east, node, [{\rulem}, node]]
\end{forest}}
\newbox\nodeboxthree
\setbox\nodeboxthree=\hbox{%
\begin{forest}
for tree={no edge},
[{\goall}, grow=east, node]
\end{forest}}
\newbox\nodeboxfour
\setbox\nodeboxfour=\hbox{%
\begin{forest}
for tree={no edge},
[{\goalm}, grow=east, node]
\end{forest}}
\begin{forest}
for tree={no edge},
shortedges [{\box\nodeboxone}, container, containerselect,
   [{\box\nodeboxtwo}, name=ruleone, container, containerselect, for children={l sep-=0.35cm,l-=0.35cm}
     [{\box\nodeboxthree}, name=goall, container, containerselect, child anchor=north]
     [{\box\nodeboxfour}, name=goalm, container, containerselect, child anchor=north]
  ]
]
\draw ([xshift=-3mm,yshift=1.9mm]root.south east)--(ruleone.north);
\draw ([xshift=-5mm,yshift=1.2mm]ruleone.south)--(goall.north);
\draw ([xshift=-5mm,yshift=1.2mm]ruleone.south east)--(goalm.north);
\end{forest}
\end{subfigure}\hfill
\begin{subfigure}[T]{0.20\textwidth}
\centering
\setbox\nodeboxone=\hbox{%
\begin{forest}
for tree={no edge},
[{\goal}, grow=east, node, l sep=1mm, [{$\dotsb$}]]
\end{forest}}
\setbox\nodeboxtwo=\hbox{%
\begin{forest}
for tree={no edge},
[{\rulel}, grow=east, node, l sep=1mm [{$\dotsb$}]]
\end{forest}}
\setbox\nodeboxthree=\hbox{%
\begin{forest}
for tree={no edge},
[{$\ncontentn{n}\mid \nnextn{n}$}, grow=east, node, l sep=1mm [{$\dotsb$}]]
\end{forest}}
\setbox\nodeboxfour=\hbox{%
\begin{forest}
for tree={no edge},
[{$\ncontentn{1}\mid \nnextn{1}$}, grow=east, node, l sep=1mm [{$\dotsb$}]]
\end{forest}}
\begin{forest}
for tree={no edge},
shortedges [{\box\nodeboxone}, name=root, container
   [{\box\nodeboxtwo}, name=containertwo, container, l sep=8mm,
     [{\box\nodeboxthree}, name=containerthree, container, child anchor=north
       [{\box\nodeboxfour}, name=containerfour, container]
     ]
  ]
]
\path (containertwo.south) -- node (dots) {$\dotsb$} (containerthree.north);
\draw ([xshift=-1mm,yshift=1.2mm]root.south)--([xshift=-1mm]containertwo.north);
\draw ([xshift=-1mm,yshift=1.2mm]containertwo.south)--([xshift=-1mm,yshift=7mm]containerthree.north);
\draw ([xshift=-1mm,yshift=-6mm]containertwo.south)--([xshift=-1mm]containerthree.north);
\draw ([xshift=-1mm,yshift=1.2mm]containerthree.south)--([xshift=-1mm]containerfour.north);
\end{forest}
\end{subfigure}
\caption{Step-wise abstraction of a search tree, from left to right.
First, the nodes' data is abstracted.
For this, node types $\vec{\alphaty}\app\noden{i}=\vec{\alphaty}\app\ncontentn{i}\times\vec{\alphaty}\app\nnextn{i}$ are introduced.
Polymorphic type arguments are omitted in the figure.
Second, for each $i$, a container type $\vec{\alphaty}\app\containern{i}$ is introduced, containing nodes of type $\vec{\alphaty}\app\noden{i}$.
These containers are drawn dotted.
Each $\vec{\alphaty}\app\nnextn{i}$ is then set to point to a follow-up container.
The last figure shows the resulting model, parametrised by a number $n$
and arbitrary container and content types for each $1\leq i\leq n$.
}\label{fig:gen-search-tree}
\end{figure}}%
We next refine the model and add support for white-box proof search.
In particular, we have to provide means to easily navigate instances of this model,
moving both inside and between containers.
Finally, we add support for extensibility in \cref{sec:zippy-extensibility}.

\subsection{Navigating Search Trees with Alternating Zippers}\label{sec:zippy-navigate}
To enable easy and composable navigation,
we use concepts from functional programming theory,
namely zippers, arrows, and monads.
We first derive a suitable concept of zippers
for a single container $\vec{\alphaty}\app\container$
and then lift it to alternating zippers for our search tree model.

Essentially, a \emph{zipper} $\vec{\alphaty}\app\zipper$,
introduced in~\cite{zipper},
is a movable focus inside a container $\vec{\alphaty}\app\container$.
It typically consists of a focused \emph{(zipper) content} $\vec{\alphaty}\app\zcontent$
and a \emph{(zipper) context} $\vec{\alphaty}\app\zctxt$.
A prototypical example is a zipper for the list type $\listty{\alphaty}$:
The content is a single element of type $\alphaty\app\zcontent\coloneqq\alphaty$,
the context stores the lists to the left and right of the focused element
$\alphaty\app\zctxt\coloneqq \listty{\alphaty}\times\listty{\alphaty}$,
and the zipper is simply the product $\alphaty\app\zipper\coloneqq \alphaty\app\zcontent\times \alphaty\app\zctxt$.
A zipper comes with a set of \emph{moves}.
Moving, for example, the list zipper $(x, (ls, r \cons rs))$
focused on $x$ to its successor $r$
results in the zipper $(r, (x \cons ls, rs))$.
Moves may not succeed,
e.g.\ when moving from the final element of a list to its non-existent successor.
Modelling moves as standard $\vec{\alphaty}\app\zipper\purefun\vec{\alphaty}\app\zipper$ functions
is hence insufficient.
A second attempt is to consider the type $\vec{\alphaty}\app\zipper\purefun\optionty{\vec{\alphaty}\app\zipper}$ instead.
But in \cref{sec:exmpl-applications},
we will discover the need for moves other than those returning options.
We require a general notion of contextual computation
(e.g.\ with failure or state) that composes.
For this, we turn to categories and arrows.

A \emph{category} is a type constructor $(\cat)$
with identity
$\id\hasty \alphaty\cat\alphaty$ and
associative composition
$(\compr)\hasty (\alphaty\cat\betaty)\purefun(\betaty\cat\gammaty)\purefun\alphaty\cat\gammaty$.
Terms of type $\alphaty\cat\betaty$ are called \emph{morphisms}.
As noted by Hughes~\cite{arrows},
categories alone are insufficient to write practical code,
e.g.\ there is no way to save inputs across computations.
Hughes hence introduced \emph{arrows},
which are categories with functions
$\arr\hasty(\alphaty\purefun\betaty)\purefun\alphaty\cat\betaty$ and
$(\arrsplit)\hasty(\alphaty_1\cat\betaty_1)\purefun(\alphaty_2\cat\betaty_2)\purefun(\alphaty_1\times \betaty_1)\cat(\alphaty_2\times \betaty_2)$
subject to some expected laws~\cite{arrows}.
Many effectful computations, e.g.\ those with failures, states, and choices,
are arrows.
We take them as the foundation for moves:
\begin{definition}[Moves]
The type of \emph{moves from $\alphaty$ to $\betaty$ (in arrow $(\cat)$)}
is defined as $\alphaty\cat\betaty$.
The type of \emph{homogenous $\alphaty$-moves (in arrow $(\cat)$)}
is defined as $\hmoveexplicit{\alphaty}{\cat}\coloneqq \alphaty\cat\alphaty$.
We drop the arrow subscript if it is clear from context.
\end{definition}
Next, we fix the set of moves that a zipper must provide.
For this, we must consider the kind of container types that we want to navigate.
In this work, we limit ourselves to datatypes whose values admit a
finitely-branching tree structure.
Following~\cite{zipper}, there are four move directions in such trees:
A $\zright$ move shifts the focus to the current element's right sibling,
and a $\zdown$  move shifts the focus to its first (from left to right)  child.
The moves $\zleft$ and $\zup$ are dual.
Moreover, we need a move $\zzip$ to initialise a zipper from a container
and its inverse $\zunzip$, re-creating a container from a zipper.
\cref{fig:zip-moves} shows an example sequence of moves.
We can now define our zipper specification:
{%
\newcommand{\assumption}{Assm}
\newcommand{\contradiction}{Contradiction}
\newcommand{\goal}{$1$}
\newcommand{\rulel}{$3$}
\newcommand{\rulem}{$2$}
\newcommand{\goalml}{$4$}
\newcommand{\goalmr}{$5$}
\begin{figure}[t]
\forestset{%
shortedges/.style={%
  for tree={l sep-=0.30cm,l-=0.30cm},
}
}
\begin{subfigure}[T]{0.135\textwidth}
\centering
\begin{forest}
tikz+={\draw [select, container draw] ([xshift=-1mm,yshift=-1mm]current bounding box.south west) rectangle ([xshift=1mm,yshift=1mm]current bounding box.north east);},
shortedges [{\goal}, node
   [{\rulem}, node
     [{\goalml}, node
     ]
     [{\goalmr}, node
     ]
   ]
   [{\rulel}, node ]
]
\end{forest}
\end{subfigure}
\begin{subfigure}[T]{0.135\textwidth}
\centering
\begin{forest}
shortedges [{\goal}, node, select
   [{\rulem}, node
     [{\goalml}, node
     ]
     [{\goalmr}, node
     ]
   ]
   [{\rulel}, node ]
]
\end{forest}
\end{subfigure}
\begin{subfigure}[T]{0.135\textwidth}
\centering
\begin{forest}
shortedges [{\goal}, node
   [{\rulem}, node, select
     [{\goalml}, node
     ]
     [{\goalmr}, node
     ]
   ]
   [{\rulel}, node ]
]
\end{forest}
\end{subfigure}
\begin{subfigure}[T]{0.135\textwidth}
\centering
\begin{forest}
shortedges [{\goal}, node
   [{\rulem}, node
     [{\goalml}, node, select
     ]
     [{\goalmr}, node
     ]
   ]
   [{\rulel}, node ]
]
\end{forest}
\end{subfigure}
\begin{subfigure}[T]{0.135\textwidth}
\centering
\begin{forest}
shortedges [{\goal}, node
   [{\rulem}, node
     [{\goalml}, node
     ]
     [{\goalmr}, node, select
     ]
   ]
   [{\rulel}, node ]
]
\end{forest}
\end{subfigure}\hfill
\begin{subfigure}[T]{0.135\textwidth}
\centering
\begin{forest}
shortedges [{\goal}, node
   [{\rulem}, node, select
     [{\goalml}, node
     ]
     [{\goalmr}, node
     ]
   ]
   [{\rulel}, node ]
]
\end{forest}
\end{subfigure}\hfill
\begin{subfigure}[T]{0.135\textwidth}
\centering
\begin{forest}
tikz+={\draw [select, container draw] ([xshift=-1mm,yshift=-1mm]current bounding box.south west) rectangle ([xshift=1mm,yshift=1mm]current bounding box.north east);},
shortedges [{\goal}, node
   [{\rulem}, node
     [{\goalml}, node
     ]
     [{\goalmr}, node
     ]
   ]
   [{\rulel}, node ]
]
\end{forest}
\end{subfigure}
\caption{Zipper navigation in a tree-shaped container,
using the move sequence $[\zzip,\zdown,\zdown,\zright,\zup,\zunzip]$.
Focused parts are bold and red, containers are dotted, and nodes are solid.
Initially, the whole container is in focus.
We move to the root using $\zzip$.
Using $\zdown$, we move to its first child.
This is once repeated.
Then we move to the node's right sibling using $\zright$.
Using $\zup$ moves to the node's parent.
Finally, $\zunzip$ moves the focus to the container.}\label{fig:zip-moves}
\end{figure}
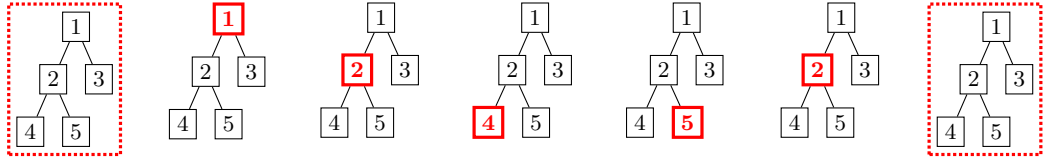}%
\begin{definition}[Zippers]
A zipper for container $\vec{\alphaty}\app\container$ in arrow $(\cat)$
is a structure $\vec{\alphaty}\app\zipper\coloneqq \vec{\alphaty}\app\zcontent\times \vec{\alphaty}\app\zctxt$
with content $\vec{\alphaty}\app\zcontent$ and context $\vec{\alphaty}\app\zctxt$
equipped with moves
$\zzip \holhasty \vec{\alphaty}\app\container \cat \vec{\alphaty}\app\zipper$,
$\zunzip \holhasty \vec{\alphaty}\app\zipper \cat \vec{\alphaty}\app\container$, and
$\zup, \zdown, \zleft, \zright\holhasty \hmove{\vec{\alphaty}\app\zipper}$.
\end{definition}
Recall that our search tree model consists of a number of alternating container types $\vec{\alphaty}\app\containern{i}$.
Based on this structure, we introduce the notions of linked and alternating zippers
that allow navigation both inside containers and between containers.
\begin{definition}[Linked Zippers]\label{def:linkedzip}
An \emph{($n$-)linked zipper} contains,
for each \makebox{$1\leq i\leq n$} and $1\leq j<n$,
a zipper $\vec{\alphaty}\app\zippern{i}$
in the fixed arrow $(\cat)$
and moves
$\zdownn{j}\holhasty \vec{\alphaty}\app\zippern{j}\cat\vec{\alphaty}\app\zippern{j+1}$ and
$\zupn{j+1}\holhasty \vec{\alphaty}\app\zippern{j + 1}\cat\vec{\alphaty}\app\zippern{j}$.
We use dot notation to access the zippers' data, e.g.\ $\zipopn{1}{\zright}$.
\end{definition}
\cref{fig:zip-moves} shows an example sequence of moves in a linked zipper.
\begin{definition}[Alternating Zippers]\label{def:altzip}
An \emph{($n$-)alternating zipper}
is an $n$-linked zipper with additional moves
$\zdownn{n}\holhasty \vec{\alphaty}\app\zippern{n}\cat\vec{\alphaty}\app\zippern{1}$ and
$\zupn{1}\holhasty \vec{\alphaty}\app\zippern{1}\cat\vec{\alphaty}\app\zippern{n}$.
\end{definition}
{%
\begin{figure}[t]
\newcommand{\goal}{$\ncontentn{1}\mid \nnextn{1}$}
\newcommand{\goall}{\goal}
\newcommand{\goalm}{\goal}
\newcommand{\rulel}{$\ncontentn{2}\mid \nnextn{2}$}
\newcommand{\rulem}{\rulel}
\begin{subfigure}[T]{0.29\textwidth}
\centering
\newbox\nodeboxone
\setbox\nodeboxone=\hbox{%
\begin{forest}
for tree={no edge},
[{\goal}, grow=east, node]
\end{forest}}
\newbox\nodeboxtwo
\setbox\nodeboxtwo=\hbox{%
\begin{forest}
for tree={edge=dashed},
[{\rulel}, grow=east, node, [{\rulem}, node]]
\end{forest}}
\begin{forest}
for tree={no edge},
shortedges [{\box\nodeboxone}, container, containerselect,
   [{\box\nodeboxtwo}, name=ruleone, container, for children={l sep-=0.35cm,l-=0.35cm}
  ]
]
\draw ([xshift=-8mm,yshift=1.1mm]root.south east)--(ruleone.north);
\end{forest}
\end{subfigure}\hfill
\begin{subfigure}[T]{0.29\textwidth}
\centering
\setbox\nodeboxone=\hbox{%
\begin{forest}
for tree={no edge},
[{\goal}, grow=east, node, select]
\end{forest}}
\setbox\nodeboxtwo=\hbox{%
\begin{forest}
for tree={edge=dashed},
[{\rulel}, grow=east, node, [{\rulem}, node]]
\end{forest}}
\begin{forest}
for tree={no edge},
shortedges [{\box\nodeboxone}, container,
   [{\box\nodeboxtwo}, name=ruleone, container, for children={l sep-=0.35cm,l-=0.35cm}
  ]
]
\draw ([xshift=-8mm,yshift=1.1mm]root.south east)--(ruleone.north);
\end{forest}
\end{subfigure}\hfill
\begin{subfigure}[T]{0.29\textwidth}
\centering
\setbox\nodeboxone=\hbox{%
\begin{forest}
for tree={no edge},
[{\goal}, grow=east, node]
\end{forest}}
\setbox\nodeboxtwo=\hbox{%
\begin{forest}
for tree={edge=dashed},
[{\rulel}, grow=east, node, select, [{\rulem}, node]]
\end{forest}}
\begin{forest}
for tree={no edge},
shortedges [{\box\nodeboxone}, container,
   [{\box\nodeboxtwo}, name=ruleone, container, for children={l sep-=0.35cm,l-=0.35cm}
  ]
]
\draw ([xshift=-8mm,yshift=1.1mm]root.south east)--(ruleone.north);
\end{forest}
\end{subfigure}\hfill

\vspace{0.5\baselineskip}
\begin{subfigure}[T]{0.29\textwidth}
\centering
\setbox\nodeboxone=\hbox{%
\begin{forest}
for tree={no edge},
[{\goal}, grow=east, node]
\end{forest}}
\setbox\nodeboxtwo=\hbox{%
\begin{forest}
for tree={edge=dashed},
[{\rulel}, grow=east, node, [{\rulem}, node, select]]
\end{forest}}
\begin{forest}
for tree={no edge},
shortedges [{\box\nodeboxone}, container,
   [{\box\nodeboxtwo}, name=ruleone, container, for children={l sep-=0.35cm,l-=0.35cm}
  ]
]
\draw ([xshift=-8mm,yshift=1.1mm]root.south east)--(ruleone.north);
\end{forest}
\end{subfigure}\hfill
\begin{subfigure}[T]{0.29\textwidth}
\centering
\setbox\nodeboxone=\hbox{%
\begin{forest}
for tree={no edge},
[{\goal}, grow=east, node, select]
\end{forest}}
\setbox\nodeboxtwo=\hbox{%
\begin{forest}
for tree={edge=dashed},
[{\rulel}, grow=east, node, [{\rulem}, node]]
\end{forest}}
\begin{forest}
for tree={no edge},
shortedges [{\box\nodeboxone}, container,
   [{\box\nodeboxtwo}, name=ruleone, container, for children={l sep-=0.35cm,l-=0.35cm}
  ]
]
\draw ([xshift=-8mm,yshift=1.1mm]root.south east)--(ruleone.north);
\end{forest}
\end{subfigure}\hfill
\begin{subfigure}[T]{0.29\textwidth}
\centering
\setbox\nodeboxone=\hbox{%
\begin{forest}
for tree={no edge},
[{\goal}, grow=east, node]
\end{forest}}
\setbox\nodeboxtwo=\hbox{%
\begin{forest}
for tree={edge=dashed},
[{\rulel}, grow=east, node, [{\rulem}, node]]
\end{forest}}
\begin{forest}
for tree={no edge},
shortedges [{\box\nodeboxone}, container, containerselect,
   [{\box\nodeboxtwo}, name=ruleone, container, for children={l sep-=0.35cm,l-=0.35cm}
  ]
]
\draw ([xshift=-8mm,yshift=1.1mm]root.south east)--(ruleone.north);
\end{forest}
\end{subfigure}\hfill
\caption{Navigation in a linked zipper, from top-left to bottom-right,
using the move sequence $[\zipopn{1}{\zzip}$, $\zdownn{1}$, $\zipopn{2}{\zright}$, $\zupn{2}$, $\zipopn{1}{\zunzip}]$.
Focused parts are bold and red, containers are dotted, and nodes are solid.
Edges inside of containers are dashed.
Initially, the container $\vec{\alphaty}\app\containern{1}$ is in focus.
Using $\zipopn{1}{\zzip}$, we move to its node.
Using $\zdownn{1}$, we move to the children's container's root.
Using $\zipopn{2}{\zright}$, we move to the node's right sibling.
Using $\zupn{2}$, we move to the container's parent.
Finally, $\zipopn{1}{\zunzip}$ moves the focus to the container.}\label{fig:azip-moves}
\end{figure}
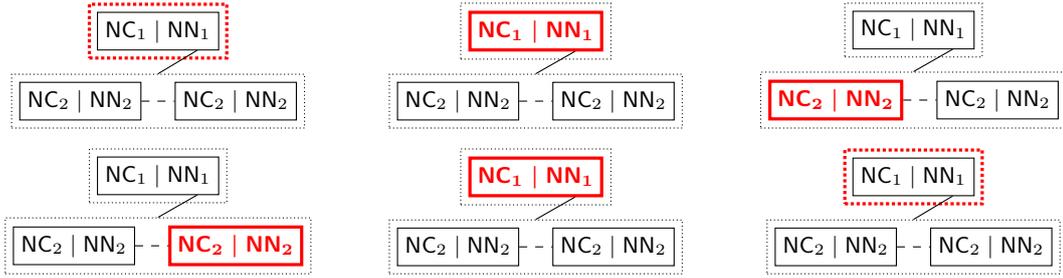}%
Alternating zippers are a specification
to navigate instances of our search tree model, thus enabling white-box searches.
A nice property of alternating zippers is that they are closed under products.
We can use this to enrich alternating zippers with
move-dependent data (such as positional information),
as we will demonstrate in \cref{sec:exmpl-applications}.
\begin{definition}[Alternating Zipper Product]\label{def:pair-altzip}
Fix two alternating zippers in the same arrow $(\cat)$ using
zippers
$\vec{\alphaty}\app\zippern{j,i}=\vec{\alphaty}\app\zcontentn{j,i}\times \vec{\alphaty}\app\zctxtn{j,i}$
for containers $\vec{\alphaty}\app\containern{j,i}$
with moves
$\zdownn{j,i}\holhasty \vec{\alphaty}\app\zippern{j,i}\cat\vec{\alphaty}\app\zippern{j,i\modplus 1}$ and
$\zupn{j,i}\holhasty \vec{\alphaty}\app\zippern{j,i}\cat\vec{\alphaty}\app\zippern{j,i\modsub 1}$
for $j\in\{1,2\}$ and $1\leq i\leq n$,
where  $\modplus,\modsub$ denote modular addition and subtraction in
$\{1,\dotsc, n\}$, i.e.\ $n\modplus 1=1$ and $1\modsub 1=n$.
The alternating zipper product is given by
\begin{align}
&\vec{\alphaty}\app\zippern{i}\coloneqq (\vec{\alphaty}\app\zcontentn{1,i}\times\vec{\alphaty}\app\zcontentn{2,i})\times (\vec{\alphaty}\app\zctxtn{1,i}\times \vec{\alphaty}\app\zctxtn{2,i}),\\
&\vec{\alphaty}\app\containern{i}\coloneqq \vec{\alphaty}\app\containern{1,i}\times \vec{\alphaty}\app\containern{2,i},
\qquad
\qquad
\zipopn{i}{move}\coloneqq \zipopn{1,i}{move}\arrsplit\zipopn{2,i}{move},\\
&\zdownn{i}\coloneqq\zdownn{1,i}\arrsplit\zdownn{2,i},
\qquad
\qquad
\zupn{i}\coloneqq\zupn{1,i}\arrsplit\zupn{2,i},
\end{align}
for $move\in\{\zzip,\zunzip,\zright,\zleft,\zup,\zdown\}$ and $1\leq i\leq n$.
\end{definition}
It would be a bit tedious to craft alternating zippers manually.
We next show how to automatically generate an important subset of them,
thus offering not just a set of specifications but a usable framework.

\subsubsection{Generating Alternating Zippers}\label{sec:gen-zippers}
We devise a mechanism to generate alternating zippers for
the common case of input zippers with single polymorphic content,
e.g.\ zippers focusing on $\alphaty$
in $\listty{\alphaty}$s or $\treety{\alphaty}$s.
We do so in two steps.
First, we show how to generate them from zippers
with node content and alternating next types,
i.e.\
$\vec{\alphaty}\app\zippern{i}=\vec{\alphaty}\app\noden{i}\times \vec{\alphaty}\app\zctxtn{i}$,
with follow-up containers stored in $\vec{\alphaty}\app\nnextn{i}$,
as is the case for our search tree model.
We call these \emph{node zippers} below.
Before we do so, we must decide how to precisely model each $\vec{\alphaty}\app\nnextn{i}$.
Nodes may have some or no children,
suggesting $\vec{\alphaty}\app\nnextn{i}=\optionty{\vec{\alphaty}\app\containern{i\modplus 1}}$.
But again, instead of picking a fixed type constructor,
we proceed with a more general solution,
leading us to monads.

A \emph{monad}~\cite{monads} is a type constructor $\monadvar$
with functions $\pure\hasty\alphaty\purefun\alphaty\app\monadvar$
and $(\bind)\hasty\alphaty\app\monadvar \purefun (\alphaty\purefun \betaty\app\monadvar)\purefun \betaty\app\monadvar$
subject to the well-known monad laws~\cite{monadprogramming}.
% In essence, monads provide a way to structure contextual computations,
% where $\pure$ lifts a value into the monad's context
% and $x \bind f$ updates the contextual value $x$ with computation $f$.
Options, lists, states, and many other computation contexts are monads.
We thus set $\vec{\alphaty}\app\nnextn{i}=\vec{\alphaty}\app\containern{i\modplus 1}\app \monadvar$
for our search tree model.
Note that doing so leaves us with two types of contextual computations in a zipper:
an arrow $(\cat)$ for moves
and a monad $\monadvar$ for $\vec{\alphaty}\app\nnextn{i}$.
In principle, there is nothing wrong with this,
but in practice, it introduces two levels of computation contexts
that have to be differentiated by the user.
For the sake of simplicity,
we opt to conflate the contexts at this stage.
Fortunately, every monad $\monadvar$ gives rise to an arrow
$\alphaty\kleisli{\monadvar}\betaty\coloneqq \alphaty\purefun\betaty\app\monadvar$,
the so-called Kleisli category~\cite{monads},
that we can use for this purpose;
that is, we require $(\cat)=(\kleisli{\monadvar})$.\footnote{%
Users requiring different computation contexts
have to provide a suitable compound monad,
for example, by using monad transformers~\cite{monadtransformer}.}
% We now describe the generation of alternating zippers for our search tree model.

\paragraph*{Generating Alternating Zippers from Node Zippers}
Fix zippers $\vec{\alphaty}\app\zippern{i}=\vec{\alphaty}\app\noden{i}\times \vec{\alphaty}\app\zctxtn{i}$
for containers $\vec{\alphaty}\app\containern{i}$
in a Kleisli category $(\kleisli{\monadvar})$
with $\vec{\alphaty}\app\noden{i}=\vec{\alphaty}\app\ncontentn{i}\times\vec{\alphaty}\app\nnextn{i}$
and $\vec{\alphaty}\app\nnextn{i}=\vec{\alphaty}\app\containern{i\modplus 1}\app \monadvar$
for $1\leq i\leq n$.
Each zipper $\zippern{i}$ stores data relevant for the navigation of
container $\containern{i}$.
We have to enrich the zippers such that they also hold
data enabling the navigation between containers.
For this, we mutually recursively define the
new zipper contexts $\vec{\alphaty}\app\anzctxtn{i}$
and types of parent data $\vec{\alphaty}\app\anzpzippern{i}$ by
\begin{equation}
\vec{\alphaty}\app\anzctxtn{i}\coloneqq \vec{\alphaty}\app\zctxtn{i}\times \vec{\alphaty}\app\anzpzippern{i}\app \monadvar
\qquad\text{and}\qquad
\vec{\alphaty}\app\anzpzippern{i}\coloneqq \vec{\alphaty}\app\ncontentn{i\modsub 1}\times \vec{\alphaty}\app\anzctxtn{i\modsub 1},
\qquad
\text{for }1 \leq i\leq n.
\end{equation}
Intuitively, we enrich the contexts with
the respective container's parent content
and, recursively, its parent's enriched context.
Next, we update the zippers to work on the enriched contexts.
The new containers are defined by
$\vec{\alphaty}\app\anzcontainern{i}\coloneqq \vec{\alphaty}\app\containern{i}\times \vec{\alphaty}\app\anzpzippern{i}\app \monadvar$
and the new zippers by
$\vec{\alphaty}\app\anzzippern{i}\coloneqq \vec{\alphaty}\app\noden{i}\times \vec{\alphaty}\app\anzctxtn{i}$.
We define the functions
$\zipperfrombasezippern{i}\holhasty \vec{\alphaty}\app\zippern{i}\times \vec{\alphaty}\app\anzpzippern{i}\app\monadvar\purefun \vec{\alphaty}\app\anzzippern{i}$ and
$\basezipperfromzippern{i}\holhasty \vec{\alphaty}\app\anzzippern{i}\purefun\vec{\alphaty}\app\zippern{i}$
by
\begin{equation}
\zipperfrombasezippern{i}\app ((zc, zx), pzm) \coloneqq
(zc, (zx, pzm))
\quad \text{and} \quad
\basezipperfromzippern{i}\app (zc, (zx, pzm)) \coloneqq
(zc, zx).
\end{equation}
Now we can lift the zippers' old moves
\begin{align}
&\opaccess{\anzzippern{i}}{\zzip}\app (co, pzm) \coloneqq
\opaccess{\zippern{i}}{\zzip}\app co \bind \arr\app (\lambda z.\app \zipperfrombasezippern{i}\app (z, pzm)),\\
&\opaccess{\anzzippern{i}}{\zunzip}\app (zc, (zx, pzm)) \coloneqq\nonumber\\
&\qquad\arr\app\basezipperfromzippern{i}\app (zc, (zx, pzm))
\bind \opaccess{\zippern{i}}{\zunzip}
\bind \arr\app (\lambda co.\app (co, pzm)).
\end{align}
For the remaining old moves, we define the lifting
$\liftmoven{i}\holhasty \hmove{\vec{\alphaty}\app\zippern{i}}\purefun \vec{\alphaty}\app\zippern{i}^{\prime\hmovesym}$
by
\begin{align}
&\liftmoven{i}\app move\app (zc, (zx, pzm))\coloneqq\nonumber\\
&\qquad\arr\app\basezipperfromzippern{i}\app (zc, (zx, pzm))
\bind move
\bind \arr\app(\lambda z.\app\zipperfrombasezippern{i}\app(z, pzm))
\end{align}
and set
$\opaccess{\anzzippern{i}}{move} \coloneqq \liftmoven{i}\app \opaccess{\zippern{i}}{move}$
for each $move\in\{\zright,\zleft,\zup,\zdown\}$.
Finally, we can define the missing moves between containers
\begin{align}
&\anzdownn{i}'\app((nc, nn), zx')\coloneqq
nn
\bind \parenths[\big]{\lambda co.\app \opaccess{\anzzippern{i\modplus 1}}{\zzip}\app (co, \pure\app (nc, zx'))},\\
&\anzupn{i}'\coloneqq
\opaccess{\anzzippern{i\modplus 1}}{\zunzip}
\compr \parenths[\big]{\lambda (co, pzm).\app pzm
\bind \arr\app\parenths[\big]{\lambda (nc, zx').\app ((nc,\pure\app co), zx')}}.
\end{align}

\paragraph*{Generating Node Zippers}
The preceding construction takes input zippers
aware of their follow-up containers,
i.e.\ $\vec{\alphaty}\app\zippern{i}=\vec{\alphaty}\app\noden{i}\times \vec{\alphaty}\app\zctxtn{i}$
and $\vec{\alphaty}\app\nnextn{i}=\vec{\alphaty}\app\containern{i\modplus 1}\app \monadvar$.
We show how to create these zippers from zippers with single polymorphic content.
Fix zippers $\alphaty\app\zippern{i}=\alphaty\times \alphaty\app\zctxtn{i}$
for containers $\alphaty\app\containern{i}$
for $1\leq i\leq n$.
We instantiate the zippers such that they operate on the required node types.
For this, we mutually recursively define the node types
\begin{align}
\alphaty^n\noden{i}\coloneqq \alphaty_i\times\alphaty^n\app\nnextn{i}
\qquad\text{and}\qquad
\alphaty^n\nnextn{i}\coloneqq
\alphaty^n\noden{i\modplus 1}\app\containern{i\modplus 1}\app \monadvar,
\qquad
\text{for }1 \leq i\leq n,
\end{align}
where $\alphaty^n\coloneqq \parenths{\alphaty_1,\dotsc,\alphaty_n}$.
The new zippers are then defined as
$\alphaty^n\app\anzzippern{i}\coloneqq\alphaty^n\noden{i}\times \alphaty^n\noden{i}\app\zctxtn{i}$
for containers
$\alphaty^n\noden{i}\app\containern{i}$
with node content $\alphaty^n\ncontentn{i}=\alphaty_i$.

\paragraph*{Summary}
To sum up, let us evaluate our progress against the desired properties from the beginning of \cref{sec:zippy-design}:
Alternating zippers support a large class of search tree structures,
thus satisfying Properties~\eqref{propty:gen-purp} and \eqref{propty:tree-based}.
They provide functionality to navigate these search trees,
thus enabling white-box searches (\proptyref{propty:white-box}).
As an abstract specification and its use of arrows and monads,
alternating zippers are independent of concrete tree representations and effects,
thus satisfying \proptyref{propty:custom}.
Finally, we provided mechanisms to generate alternating zippers from simple input zippers.
We next turn to the missing Property~\eqref{propty:extensible} -- extensibility.

\subsection{Extensible Search Trees}\label{sec:zippy-extensibility}
As discussed in \cref{sec:intro},
state-of-the-art white-box proof search tools offer little to no extensibility.
One reason is that they are implemented against specifications with little room for variation.
Alternating zippers abstract away from search tree-specific data.
Generic code for alternating zippers remains usable
even if one's concrete search tree's data representation or nodes' content changes.
However, alternating zippers only provide a specification for search tree navigation.
Data manipulation, i.e.\ viewing and updating the search tree's and its nodes' content, is not covered.
Naturally, we cannot write sensible proof automation without modifying any of the search tree's data.
But as it is, any such code will be coupled to the specific search tree's data representations,
which goes against our goal of extensibility.
Just like we derived a specification for search tree navigation,
we now need a specification framework to talk about data manipulation.
Luckily, programming theory offers a solution to our problem: lenses~\cite{lensorig,lenses}.

Essentially, a lens for $\alphaty,\betaty$ describes how to \emph{get}
a value of type $\betaty$ from a value of type $\alphaty$
and how to \emph{modify} a value of type $\betaty$ inside of $\alphaty$.
Formally, a \emph{lens in Kleisli category $(\kleisli{\monadvar})$}
is a value of type
$\lens{\alphaty}{\betaty}_\monadvar\coloneqq \getter{\alphaty}{\betaty}_\monadvar\times \modifier{\alphaty}{\betaty}_\monadvar$,
where
$\getter{\alphaty}{\betaty}_\monadvar\coloneqq \alphaty\kleisli{\monadvar}\betaty$
and
$\modifier{\alphaty}{\betaty}_\monadvar\coloneqq (\betaty\kleisli{\monadvar}\betaty)\times\alphaty\kleisli{\monadvar}\alphaty$.
We drop the $\monadvar$ subscript if clear from context.
Lenses are not only a framework to talk about but also simplify data manipulation.
The reason is that they happily compose, forming a category.
For the curious:
\begin{align}
\id&\coloneqq\parenths[\big]{\pure, \lambda (f, x).\app f\app x},\\
(g_1,m_1) \compr (g_2,m_2)&\coloneqq\parenths[\big]{g_1\compr_\monadvar g_2,
\lambda (f, x).\app m_1\app (\lambda y.\app m_2\app (f, y), x)},
\end{align}
where $\compr_\monadvar$ is Kleisli composition.
We can thus create lenses for complexly nested data
by composition of simpler sublenses.
\begin{example}
Consider a node zipper $\zipper=\node\times \zctxt$
with $\node=\ncontent\times\nnext$.
Given a lens
$l_{\zipper,\node} \holhasty \lens{\zipper}{\node}$
for the zipper's node
and a lens
$l_{\node,\ncontent}\holhasty \lens{\node}{\ncontent}$,
for the node's content,
we can create the lens
$l_{\zipper,\ncontent}\holhasty \lens{\zipper}{\ncontent}$
for the zipper's node content by
$l_{\zipper,\ncontent}\coloneqq l_{\zipper,\node}\compr l_{\node,\ncontent}$.
\end{example}
Combining alternating zippers with lenses gives us a specification framework
to navigate search trees and manipulate their data
independent of concrete data representations.
Let us concretise this by means of an example.
{%
\newcommand{\zerochild}{\constfont{zerochild}}
\newcommand{\genzerochild}{\zerochild'}
\begin{example}\label{exmpl:lens}
Assume we have a 2-alternating zipper
in Kleisli category $(\kleisli{\monadvar})$
with zipper content $\zcontentn{2}= \natty$.
We can define a function
$\zerochild\holhasty \zippern{1}\kleisli{\monadvar} \zippern{2}$
to zero the focused node's first child's number by
\begin{equation}
\zerochild\coloneqq
\zdownn{1}\compr\arr\parenths[\big]{\lambda (n, zx).\app (0, zx)}.
\end{equation}
As it is, $\zerochild$ is tied to the concrete representation of
$\zcontentn{2}$.
If we, for example, extend the second zipper's content with a second field,
say $\zcontentn{2}= (\natty, \stringty)$,
the function $\zerochild$ is of no use.
To keep our development extensible, we can rewrite the code using a lens
$(get, modify)\holhasty \lens{\zcontentn{2}}{\natty}_\monadvar$.\footnote{In ML-languages, such as \isabelleml, the lens may be passed as a ML-functor argument.}
We define the function $\genzerochild\holhasty \zippern{1}\kleisli{\monadvar}\zippern{2}$
as follows
\begin{equation}
\genzerochild\coloneqq
\zdownn{1}\compr\parenths[\big]{\lambda (zc, zx).\app modify\app (\arr\app (\lambda \underscorematch.\app 0), zc)\bind\arr\app(\lambda zc.\app (zc, zx)}.
\end{equation}
This way, the function remains usable independent of changes to the second zipper's content.
Certainly, this involved some overhead for the function in question.
But the benefits start to shine when we deal with larger architectures,
such as complex proof search automation.
\end{example}}%
As a final step, we add a design principle
for data extension of search tree instances.
Our approach is inspired by \isabellehol's
extensible record schemes~\cite{recordmore}.
The idea is to represent data slots, e.g.\ zipper contents,
as records containing an extra polymorphic $more$ field.
Whenever we wish to extend a data slot,
we instantiate its $more$ field with a record containing the new data and a new $more$ field.
Again, let us concretise this with an example.
\begin{example}
Consider an $n$-alternating zipper
with zippers
$\zippern{i}=\zcontentn{i}\times \zctxtn{i}$
for containers $\containern{i}$
with content
$\zcontentn{i}=\natty$
for $1\leq i\leq n$.
As it is, there is no general way
to extend the alternating zipper such that,
for example, $\zcontentn{1}$ additionally contains a $\stringty$ value.
Assume now that the zippers are of the form
$\alphaty^n\zippern{i}=\alphaty^n\app\zcontentn{i}\times \alphaty^n\zctxtn{i}$
for containers
$\alphaty^n\containern{i}$
with contents
$\alphaty^n\app\zcontentn{i}= (\natty\times \alphaty_i)$.
The polymorphic variables $\alphaty_i$ are used as the zippers' $more$ fields.
In this case, we can extend $\zippern{1}$'s content with a $\stringty$ value
by instantiation:
We
define $\alpha^n_{inst}\coloneqq (\stringty\times\alphaty_1,\alphaty_2,\dotsc,\alphaty_n)$
and set
$\alphaty^n\zippern{i}'\coloneqq \alpha^n_{inst}\app\zippern{i}$,
obtaining zippers for the containers
$\alpha^n_{inst}\containern{i}$.
\end{example}
All in all, the framework achieves extensibility
due to abstract specifications for search tree navigation and data manipulation,
as well as a data extension principle for search tree instances.
This concludes the design of \Zippy.

\section{Implementing Zippy in \isabelleml}\label{sec:zippy-impl}
In this section, we describe a common pitfall
when implementing monads, arrows, lenses, and the like with ML's module system
and how we take care of it in \isabelleml.
Then we sketch how $n$-alternating zippers
can be defined using metaprogramming for \isabelleml.
For an introduction to ML's module system, we refer to Paulson's book~\cite[Chapter 7]{mlbook}.

\paragraph*{Monads, Arrows, Lenses in \isabelleml}
It is known that monads and other type constructor specifications can be modelled with ML's module system,
exemplified by various blog posts online\footnote{%
e.g.\ by Robert Harper \url{https://web.archive.org/web/20250214092012/https://existentialtype.wordpress.com/2011/05/01/of-course-ml-has-monads/}}
and a specification of monads in \isabelleml by the \psl framework~\cite{pslmonads}.
However, expositions for serious applications are largely missing.
Indeed, the standard approach entails a pitfall,
which we demonstrate with an example:
\begin{example}
The \emph{state monad} $\state{\statetyvar}{\alphaty}\coloneqq\statetyvar\purefun \alphaty\times\statetyvar$
takes a state of type $\statetyvar$
and produces a value of type $\alphaty$ along with an updated state.
It is a monad with operations
$\pure\app x\app s\coloneqq (x,s)$
and
$(m\bind\app f)\app s\coloneqq (\lambda (x, s').\app f\app x\app s')\app(m\app s)$.
To implement state monads in ML-languages, we first create the signature of monads:
\begin{lstlisting}[language=myml]
signature MONAD =
sig
  type 'a t
  val pure : 'a -> 'a t
  val bind : 'a t -> ('a -> 'b t) -> 'b t
end
\end{lstlisting}
Next, we implement the state monad.
\begin{lstlisting}[language=myml]
structure State =
struct
  type ('s, 'a) t = 's -> 'a * 's
  fun pure x s = (x, s)
  fun bind m f s = let val (x, s') = m s in f x s' end
end
\end{lstlisting}
Unfortunately, \inlineml{State} cannot be made an instance of \inlineml{MONAD}.
The issue is that \inlineml{MONAD} expects a unary type constructor
\inlineml{'a t}
while \inlineml{State} uses a binary one, \inlineml{('s, 'a) t},
to accommodate for arbitrary states \inlineml{'s}.
A common solution is to make \inlineml{State} an ML-functor:
\begin{lstlisting}[language=myml]
functor State'(type s) : MONAD =
struct
  type 'a t = s -> 'a * s
  fun pure x s = (x, s)
  ...
\end{lstlisting}
But mainstream ML-languages, particularly Standard ML,
do not allow polymorphic functor instances such as
\inlineml{State'(type s = 's)}.
As a result, users have to create separate instances of \inlineml{State'}
for every concrete state type they use in their code.
To obtain a state-polymorphic monad instance of \inlineml{State} in \isabelleml,
there is no option but to generalise the signature of \inlineml{MONAD}
to include a polymorphic parameter \inlineml{'p1}
that is passed through the monad's operations:
\begin{lstlisting}[language=myml]
signature MONAD_1 =
sig
  type ('p1, 'a) t
  val pure : 'a -> ('p1, 'a) t
  val bind : ('p1, 'a) t -> ('a -> ('p1, 'b) t) -> ('p1, 'b) t
end
\end{lstlisting}
Now it indeed holds that \inlineml{State : MONAD_1}.
Moreover, every \inlineml{M : MONAD} is a \inlineml{MONAD_1} with
type \inlineml{('p1, 'a) t = 'a M.t}.
\end{example}
The example highlights a general problem:
$(k+1)$-ary type constructors \inlineml{('p1,...,'pk,'a) t}
monadic in \inlineml{'a}
require corresponding $(k+1)$-ary types in a \inlineml{MONAD_k} signature.
It would be tedious to create these manually for each $k$.
We use metaprogramming for \isabelleml by means of its antiquotation mechanism~\cite{isabelleml}
to write arity-independent code that can be instantiated for arbitrary $k$.
We use antiquotations
\lstinline[mathescape]{$\text{\pargs{ts}}$},
printing $k$ polymorphic types \inlineml{'p1,...,'pk}
along with its passed type arguments \inlineml{ts},
and \lstinline[mathescape]{$\text{\npargs}$},
printing the number $k$ that is configurable as Isabelle context data.
The generic monad signature can then be implemented as follows:
\begin{lstlisting}[language=myml]
signature MONAD_`\npargs` =
sig
  type `\pargs{'a}` t
  val pure : `\pargs{'a}` -> `\pargs{'a}` t
  val bind : `\pargs{'a}` t -> ('a -> `\pargs{'b}` t) -> `\pargs{'b}` t
end
\end{lstlisting}
This approach can be extended to other common type constructor classes.
We created a library including
applicatives, monads, monad transformers, traversables, arrows, lenses,
and various instances, such as Kleisli categories, and state and list monads
in \isabelleml.
% Aesop implemented mutable references
% StandardML signatures, structures, functors.

\paragraph*{Alternating Zippers in \isabelleml}
\Zippy is centrally built around the concept of $n$-alternating zippers (\cref{def:altzip}),
which contain zippers $\vec{\alphaty}\app\zippern{i}$
for \makebox{$1\leq i\leq n$}
in an arrow $(\cat)$.
As such,
their definition in \isabelleml must be parametric in the number of type arguments $\vec{\alphaty}$,
the number of zippers $n$,
and the number of polymorphic parameters for the underlying arrow
\lstinline[mathescape]{$\text{\npargs}$}.
We use four additional antiquotations to obtain generic code:
(1) \lstinline[mathescape]{$\text{\eval{t}}$}
to print \inlineml{t}'s integer result,
(2) \lstinline[mathescape]{$\text{\zargs}$}
to print the list of polymorphic arguments $\vec{\alphaty}$,
(3) \lstinline[mathescape]{$\text{\znargs}$}
to print the number $|\vec{\alphaty}|$, and
(4) \lstinline[mathescape]{$\text{\imap{\{i\} => t}}$}
to print $t[j/\{i\}]$ for $1 \leq j\leq n$,
where $t[j / \{i\}]$
denotes syntactic substitution of any occurrence of $\{i\}$ by $j$ in $t$.
We only show an excerpt to give an impression of how the code looks like --
the details are technical and can be found in the
implementation in the supplementary material.
\begin{lstlisting}[language=myml,mathescape]
signature `\eval{$n$}`_ALTERNATING_ZIPPER_`\npargs`_`\znargs` =
sig
  include ARROW_`\npargs`
  `\imap{\{i\} => structure Z\{i\} :\ ZIPPER\_\npargs\_\znargs}`
  `{\color{mlantiquot}\antiquot{imap}\cartoucheleft{\{i\} => val down\{i\} :}}`
    `\color{mlantiquot}\pargs{\zargs\ Z\{i\}.zipper, \zargs\ Z\eval{$\{i\}\modplus 1$}.zipper} cat\cartoucheight`
...
\end{lstlisting}
We note that users of the framework are not forced to use these antiquotations
but can write explicit \isabelleml code for a fixed configuration of their liking.

\section{Proof Search with \Zippy in Isabelle}\label{sec:exmpl-applications}

We next demonstrate one way of building a concrete proof search tool with \Zippy in Isabelle.
We provide liftings to embed Isabelle tactics as search steps
and demonstrate how to implement a best-first proof search akin to Lean's \aesop.
We first devise an abstract base model for the tool and then
highlight some implementation steps and extensions.
The presented tool is kept simple for the sake of exposition.
The supplementary material contains the complete code
for an extended version of the model presented here.

\paragraph*{Goal States, Meta Variables, and Tactics in Isabelle}

An Isabelle \emph{goal state} consists of a list of goals (i.e.\ theorems to be proven)
$[G_1,\dotsc,G_n]$.
The goals may contain and share meta variables (also called schematic variables) $\mvar{x}$.
A meta variable can be instantiated with a term.
For soundness, instantiations of meta variables must be consistent across all goals.
For example,
the goal state
$[\mvar{P}\land \mvar{Q}$, $\mvar{Q}]$
may be instantiated to $[P\land \top$, $\top]$
but not $[P\land \top$, $\mvar{Q}]$.

An Isabelle \emph{tactic} is a function taking a goal state
and returning a lazy sequence of successor goal states.
The sequence models alternatives, not conjunctions.
For example,
a tactic for disjunction introduction may map a goal state $P\lor Q$
to the goal state sequence $\brackets[\big]{[P], [Q]}$
while a tactic for conjunction introduction may map a goal state $P\land Q$
to the sequence $\brackets[\big]{[P, Q]}$.
Many tactics in Isabelle are goal-indexed,
meaning that they operate on a chosen goal $G_i$ of the goal state.
Such tactics take the goal index $i$ as an extra argument.
Since tactics are arbitrary functions,
they are black-box,
offering no structure (besides that they return alternatives) to analyse their proof search.
But of course
they can still be used as proof steps to expand the search tree in
white-box proof search tools like ours.
Indeed, virtually all automation in Isabelle is mapped to a tactic at the end of the day,
so it is essential to enable their usage in our tool.

\paragraph*{The Basic Proof Search Tree Model}

We next devise the base structure of the search tree model we will implement with \Zippy.
As a starting point, we take the model presented in \cref{fig:aesop-tree-basic},
containing goal and prioritised rule nodes,
and extend it to enable the embedding of Isabelle's goal states and tactics.

First, we generalise goal nodes to contain an Isabelle goal state instead of a single goal.
Second, we add goal indices to rule nodes to indicate a rule's goal focus.
Then we introduce \emph{goal clusters} to reduce the number of duplicate proof step computations:
Consider, for example, the goal state $[A\land B, C\land D]$.
A tactic for conjunction introduction may independently be applied to goal 1 and goal 2 in any order.
An exhaustive search tree, depicted in \cref{fig:no-split-goalclusters},
tries each order, resulting in equivalent successor states.
As a remedy, we split goal states into independent goal clusters,
resulting in search trees as depicted in \cref{fig:split-goalclusters}.
However, we may not just put every goal into a separate cluster.
The reason is that we have to consistently instantiate meta variables during the search.
Our goal clusters will hence collect goals that transitively share a meta variable.
These clusters are called \emph{meta variable clusters} in Lean's \aesop~\cite{aesop}.
Formally, goal clusters are the transitive closure of the relation
denoting whether two goals share a meta variable.
Goal clusters form an equivalence relation.
For example, the goal clusters of goal state
$[\mvar{x}\land \mvar{y}$, $\mvar{v}$, $\mvar{y}\land \mvar{z}$, $\mvar{z}]$
are
$[\mvar{x}\land \mvar{y}, \mvar{y}\land \mvar{z}, \mvar{z}]$
and
$[\mvar{v}]$.

{%
\newcommand{\goal}{$A\land B, C\land D$}
\newcommand{\assumption}{Assm}
\newcommand{\contradiction}{Contradiction}
\newcommand{\rulel}{${\land}I$, 50\%, 1}
\newcommand{\rulem}{${\land}I$, 40\% 2}
\newcommand{\goall}{$A,B,C\land D$}
\newcommand{\rulell}{${\land}I$, 30\% 3}
\newcommand{\goalm}{$A\land B, C, D$}
\newcommand{\rulemm}{${\land}I$, 20\% 1}
\newcommand{\goalll}{$A,B,C,D$}
\newcommand{\goalml}{$A,B,C,D$}
\forestset{%
shortedges/.style={%
  for tree={l sep-=0.30cm,l-=0.30cm},
}
}
\begin{figure}[t]
\begin{subfigure}[T]{0.30\textwidth}
\centering
\begin{forest}
shortedges [{\goal},  goal
  [{\rulel}, rule, select]
  [{\rulem}, rule]
]
\end{forest}
\end{subfigure}\hfill
\begin{subfigure}[T]{0.30\textwidth}
\centering
\begin{forest}
shortedges [{\goal},  goal
   [{\rulel}, rule [
     {\goall}, goal
       [{\rulell}, rule]]
   ]
   [{\rulem}, rule, select]
 ]
\end{forest}
\end{subfigure}\hfill
\begin{subfigure}[T]{0.4\textwidth}
\centering
\begin{forest}
shortedges [{\goal},  goal
   [{\rulel}, rule [
     {\goall}, goal
       [{\rulell}, rule, select]]
   ]
   [{\rulem}, rule [
     {\goalm}, goal
       [{\rulemm}, rule]]
   ]
 ]
\end{forest}
\end{subfigure}\hfill

\vspace{0.5\baselineskip}
\begin{subfigure}[T]{0.5\textwidth}
\centering
\begin{forest}
shortedges [{\goal},  goal
   [{\rulel}, rule [
     {\goall}, goal
       [{\rulell}, rule
         [{\goalll}, goal
         ]
       ]]
   ]
   [{\rulem}, rule [
     {\goalm}, goal
       [{\rulemm}, rule, select]]
   ]
 ]
\end{forest}
\end{subfigure}\hfill
\begin{subfigure}[T]{0.5\textwidth}
\centering
\begin{forest}
shortedges [{\goal},  goal
   [{\rulel}, rule [
     {\goall}, goal
       [{\rulell}, rule
         [{\goalll}, goal, proof
         ]
       ]]
   ]
   [{\rulem}, rule [
     {\goalm}, goal
       [{\rulemm}, rule
         [{\goalml}, goal, proof
         ]
       ]]
   ]
]
\end{forest}
\end{subfigure}
\caption{Proof search without goal clusters. Goal nodes are solid, and rule nodes are dashed, containing a rule, priority, and goal index. Selected nodes are bold and red. The duplication is marked in bold and blue.}\label{fig:no-split-goalclusters}
\end{figure}
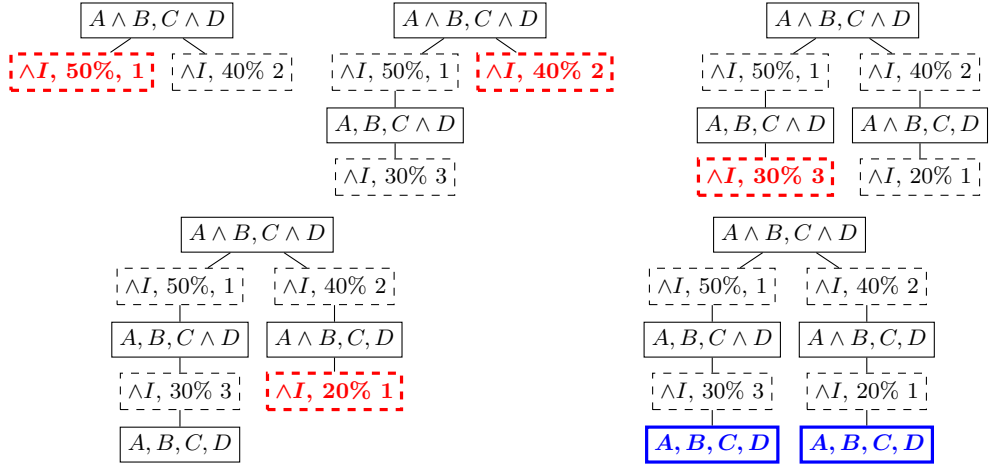
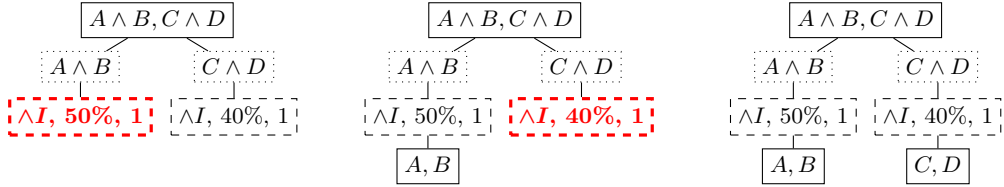
\begin{figure}[t]
\newcommand{\goalsl}{$A\land B$}
\renewcommand{\rulem}{${\land}I$, 40\%, 1}
\newcommand{\goalsr}{$C\land D$}
\renewcommand{\goall}{$A,B$}
\renewcommand{\goalm}{$C,D$}
\forestset{%
shortedges/.style={%
  for tree={l sep-=0.35cm,l-=0.35cm},
}
}
\begin{subfigure}[T]{0.33\textwidth}
\centering
\begin{forest}
shortedges [{\goal},  goal
  [{\goalsl}, goalcluster
    [{\rulel}, rule, select]
  ]
  [{\goalsr}, goalcluster
    [{\rulem}, rule]
  ]
]
\end{forest}
\end{subfigure}\hfill
\begin{subfigure}[T]{0.33\textwidth}
\centering
\begin{forest}
shortedges [{\goal},  goal
  [{\goalsl}, goalcluster
    [{\rulel}, rule [
      {\goall}, goal
    ]
  ]]
  [{\goalsr}, goalcluster
     [{\rulem}, rule, select]
  ]
 ]
\end{forest}
\end{subfigure}\hfill
\begin{subfigure}[T]{0.33\textwidth}
\centering
\begin{forest}
shortedges [{\goal},  goal
   [{\goalsl}, goalcluster
    [{\rulel}, rule [
      {\goall}, goal
    ]
  ]]
  [{\goalsr}, goalcluster
    [{\rulem}, rule [
       {\goalm}, goal
      ]
    ]
  ]
 ]
\end{forest}
\end{subfigure}\hfill
\caption{Proof search with goal clusters. Goal nodes are solid, goal clusters are dotted, and rule nodes are dashed.}\label{fig:split-goalclusters}
\end{figure}}%

Next, we generalise rule nodes to nodes containing actions.
An \emph{action} is a morphism in the zipper's arrow
that takes the alternating zipper at the action's node position and returns an updated zipper.
One important kind of action is a \emph{tactic action}.
Tactic actions apply a tactic to the action node's parent goal cluster
and attach the resulting successor states as children.
Actions can be selected multiple times.
For example, a tactic action may choose to attach its tactic's results one at a time.

Finally, we add \emph{action application nodes} beneath action nodes,
storing metadata linked to the creation of a successor goal state by an action.
For example, actions can mark their results as ``promising'',
indicating that they should be returned to the user even if proof search fails to solve some of its subgoals.
This ensures that users receive useful feedback even when a proof search attempt is incomplete.

To sum up, we use four node types and hence build our tool on top of a 4-alternating zipper:
(1) \emph{goal state nodes} store an Isabelle goal state,
(2) \emph{goal cluster nodes} store one of their parent's goal clusters,
(3) \emph{action nodes} store actions and goal indices, and
(4) \emph{action application nodes} store metadata created by an action's application.
We take four list zippers
and generate the base alternating zipper as described in \cref{sec:gen-zippers},
resulting in
$\alphaty^4\app\zippern{i}=\alphaty^4\app\noden{i}\times \alphaty^4\app\zctxtn{i}$
for containers $\alphaty^4\app\containern{i}=\listty{\alphaty_i}$
with $\alphaty^4\app\noden{i}=\alphaty_i\times\alphaty^4\app\nnextn{i}$
and $\alphaty^4\app\nnextn{i}=\alphaty^4\app\containern{i\modplus 1}\app \monadvar$
for $1\leq i\leq 4$.
We keep the Kleisli category $(\kleisli{\monadvar})$ abstract for now.

\paragraph*{Adding Goal Clusters}

When initialising a search tree with an Isabelle goal state $[G_1,\dotsc,G_n]$,
we have to construct the goal state's equivalence relation of goal clusters.
We achieve this with an imperative union-find structure that we ported to \isabelleml.
It is indeed the only part of our implementation using imperative features.
We use a term index mapping each meta variable to its (initially empty) equivalence class of goals.
For each goal, we take its meta variables and
imperatively merge the goal's equivalence class
with the class indexed by the term index.

\paragraph*{Adding Actions}
To add actions to the search tree,
we extend the alternating zipper's data as described in~\cref{sec:zippy-extensibility}.
The type of actions should be of the form $\alphaty^4\app\action\coloneqq \hmove{\alphaty^4\app\zippern{3}}$,
i.e.\ an arbitrary morphism updating the alternating zipper at its selected action node.
Note, however, that this type should talk about zippers that already include such actions.
We thus need to define the type of actions recursively:
\begin{equation}
\alphaty^4\app\action\coloneqq \alphaty^4\app \zippern{3}^{\prime\hmovesym}
\qquad \text{and} \qquad
\alphaty^4\app\zippern{i}'\coloneqq (\alphaty_1,\alphaty_2,\alphaty^4\app\action\times \alphaty_3,\alphaty_4)\app\zippern{i}.
\end{equation}
We proceed analogously to add the remaining data we wish for in action nodes.
For example, to add priorities, we define
$\alphaty^4\app\zippern{i}''\coloneqq (\alphaty_1,\alphaty_2,\typefont{prio}\times \alphaty_3,\alphaty_4)\app\zippern{i}'$,
for some desired priority type $\typefont{prio}$.
Of course, in the actual implementation,
each extension is accompanied by an appropriate specification (i.e.\ ML signature) that
abstracts from the concrete representation with the help of lenses.

As discussed, using Isabelle tactics as actions is quite essential.
We sketch one way to do this next.
This is meant to give the reader an impression of the complexity
of adding slightly more difficult functionality.
For ease of notation, we assume that tactics return standard lists, not lazy sequences.
The function
$\tacaction\holhasty \tacticty\purefun \alphaty^4\app\action$
is defined by
{%
\newcommand{\getgcstate}{\constfont{getGCState}}
\newcommand{\addaanode}{\constfont{addAANode}}
\newcommand{\initgoal}{\constfont{initGS}}
\newcommand{\listaction}{\constfont{listAction}}
\newcommand{\setaction}{\constfont{setAction}}
\newcommand{\setpriority}{\constfont{setPriority}}
\newcommand{\mkAA}{\constfont{mkAANode}}
\newcommand{\modifynext}{\constfont{modifyNext}}
\begin{align}
&\tacaction\app tac\app z\coloneqq
\zupn{3}\app z\bind
\getgcstate\bind
\arr\app tac\bind
(\lambda ss.\app \listaction\app ss\app z)\\
&\listaction\app []\app z\coloneqq \setpriority\app\constfont{Prio.disabled}\app z\\
&\listaction\app (s \cons ss)\app z \coloneqq
\addaanode\app s\app z\bind
\zupn{4}\bind \setaction\app (\listaction\app ss)\\
&\addaanode\app s\app z\coloneqq
\app \zipopn{3}{\modifynext}\app ((\lambda ns.\app \mkAA\app s\cons ns), z)
\bind \zdownn{3}
\bind \initgoal\app s
\end{align}
where $\getgcstate$ is the lens getter for a goal cluster's state,
$\setpriority,\setaction,\modifynext$ the obvious lens setters/modifiers,
$\mkAA$
a function creating a new node $\noden{4}$ from the tactic's result,
and $\initgoal$
a morphism initialising and attaching the goal and goal clusters
described by the result.
The function $\listaction$
attaches its passed results as children, one at a time,
and disables itself once the results are empty.
}

\paragraph*{Adding Failure and State}

Thus far, we kept the zippers' Kleisli category $(\kleisli{\monadvar})$ abstract.
One computation effect we have to support is failure,
e.g.\ for moves beyond the end of a zipper's container.
We implemented an option monad for this purpose.
Another helpful effect is state:
Many tactics and functions in \isabelleml require access to the so-called \emph{proof context}.
The proof context stores various data,
such as registered theorems used by the simplifier, term parsing options,
and fixed variables.
Typically, users explicitly pass around the context and use it wherever needed.
We can make this context passing implicit using a monad supporting state,
which we also implemented.
Of course, the monad's state is polymorphic, as discussed in \cref{sec:zippy-impl},
and can thus be used to store arbitrary data, not just proof contexts.
We use monad transformers to combine the failure and state effects in the implementation.

\paragraph*{Adding Positional Information}

It is useful to know the zipper's current location in the search tree.
For example, one can use positional information to limit the proof search depth,
to adjust action priorities based on their depth in the search tree,
to implement a replay mechanism for movements,
and it serves as a debugging aid when inspecting proof searches.
Adding positional information with \Zippy is strikingly simple:
we just create an alternating zipper for positional information
and pair it with the search tree's alternating zipper.

We can model a zipper's position as an integer list,
where the list's length encodes the (vertical) depth and the list's elements
the horizontal position at each level of depth.
For example, the position $[2,0,-1]$  (read it right to left) corresponds to the sequence of moves
$[\zleft,\zdown,\zdown,\zright,\zright]$.
An alternating zipper's position is just a list of zipper positions,
i.e.\ $\listty{\listty{\intty}}$.
An $n$-alternating zipper for positions can be implemented with
zippers $\zippern{i}\coloneqq\listty{\listty{\intty}}\times ()$
for containers $\containern{i}\coloneqq \listty{\listty{\intty}}$
and moves
\begin{align}
&\zipopn{i}{\zzip}\app pss\coloneqq \pure\app ([0] \cons pss),
\qquad
\qquad
\qquad
\quad
\zipopn{i}{\zunzip}\app (ps\cons pss)\coloneqq \pure\app pss,\\
&\zipopn{i}{\zright}\app ((p\cons ps)\cons pss)\coloneqq \pure\app ((p + 1\cons ps)\cons pss),\\
&\zipopn{i}{\zleft}\app((p\cons ps)\cons pss)\coloneqq \pure\app ((p - 1\cons ps)\cons pss),\\
&\zipopn{i}{\zdown}\app (ps\cons pss)\coloneqq \pure\app((0 \cons ps)\cons pss),
\quad
\zipopn{i}{\zup}\app ((p\cons ps)\cons pss)\coloneqq \pure\app (ps\cons pss),\\
&\zdownn{i}\app pss\coloneqq \pure\app ([0] \cons pss),
\qquad
\qquad
\qquad
\quad
\zupn{i}\app (ps \cons pss)\coloneqq \pure\app pss,
\end{align}
for \makebox{$1\leq i\leq n$}.
Using \cref{def:pair-altzip},
we can adjoin this alternating zipper to the one we created for the proof search tree,
giving us positional information for free.
This approach can be used to add also other move-dependent data,
such as an explicit history of moves that may be used for debugging and algorithmic analyses.

\paragraph*{Running a Best-First Search}
Finally, we sketch the implementation of a best-first search akin to Lean's \aesop.
The best-first search picks the tree's highest priority action, runs the action, and then repeats.
To pick the highest-scoring action, we need a way to visit every action node of an alternating zipper.
This, in turn, requires a way to visit the nodes of each zipper,
e.g.\ by means of two enumeration morphisms
$\zipopn{i}{\enumfirst}\holhasty \containern{i}\cat \zippern{i}$
and
$\zipopn{i}{\enumnext}\holhasty \hmove{\zippern{i}}$.
We assume that our Kleisli category ($\kleisli{\monadvar}$) supports failures,
providing a function
$\arrcatch\holhasty (\alphaty \cat \betaty) \purefun (\alphaty \cat \betaty) \purefun (\alphaty \cat \betaty)$
such that $\arrcatch\app a_1\app a_2$
is an arrow running $a_2$ in case $a_1$ failed.
A generic postorder depth-first enumeration, for example, can then be implemented by
\begin{align}
&\zipopn{i}{\enumfirst}\coloneqq \zipopn{i}{\zzip}\compr\arrrepeat\app \zipopn{i}{\zdown},\\
&\zipopn{i}{\enumnext}\coloneqq \arrcatch\app(\zipopn{i}{\zright}\compr \arrrepeat\app \zipopn{i}{\zdown})\app \zipopn{i}{\zup},\\
&\arrrepeat\app move\app z \coloneqq \arrtry\app (move \compr \arrrepeat\app move)\app z,
\qquad
\arrtry\app move \coloneqq \arrcatch\app move\app \id,
\end{align}
assuming that
$\zipopn{i}{\zzip}$
moves to container $\containern{i}$'s top-left node and
$\zipopn{i}{\zdown}$
to the leftmost child.
Using the zippers' enumerations,
it is a routine exercise to write morphisms
$\enumfirst\holhasty \containern{1}\cat \zippern{3}$
and
$\enumnext\holhasty \hmove{\zippern{3}}$
enumerating all action nodes of our proof search zipper.
We then implement a morphism
$\maxaction\holhasty \containern{1}\cat \zippern{3}$
that folds this enumeration,
returning the zipper containing the highest-scoring action.
This action can then be applied and the process repeated:
{%
\newcommand{\getaction}{\constfont{getAction}}
\begin{align}
\bestfirst&\coloneqq\arrrepeat\app (\maxaction\compr \applyaction\compr\totop)\\
\applyaction\app z&\coloneqq\getaction\app z\bind(\lambda action.\app action\app z)\\\
\totop&\coloneqq\arrrepeat\app (\zupn{3}\compr \zupn{2}\compr \zupn{1}\compr \zupn{4})
\compr \zupn{3}\compr \zupn{2}\compr \zipopn{1}{\zunzip}
\end{align}}%
Naturally, this implementation of $\bestfirst$
is kept simple for the sake of exposition,
e.g.\ it loops in case indefinitely many steps can be applied.

% {%
% \newcommand{\nextnoupn}[1]{\constfont{nextDown}_{#1}}
% \begin{align}
% &\enumfirst_{i}\app z\coloneqq (\zipopn{i}{\enumfirst} \compr \nextnoupn{i})\app z\\
% &\nextnoupn{1} \app z \coloneqq \arrtry\app (\zipopn{1}{\zdown}
% \compr \zipopn{1\modplus 1}{\zunzip}
% \compr \enumfirst_{1\modplus 1})\app z\\
% &\nextnoupn{i} \app z \coloneqq \arrtrystep\app (\zipopn{i}{\zdown}
% \compr \zipopn{i\modplus 1}{\zunzip}
% \compr \enumfirst_{i\modplus 1})\app \zipopn{i}{\enumnext}\app z\\
% &\arrtrystep\app f\app step\app z\coloneqq \arrcatch\app f\app(step\compr \arrtrystep\app f\app step)\app z\\
% &\enumnext \app z \coloneqq
% (\arrcatch\app (\zipopn{1}{\enumnext}\compr \nextnoupn{1})\\
% &\qquad(\zipopn{1}{\zup}\compr \arrcatch\app (\zipopn{n}{\enumnext}\compr \nextnoupn{n})\\
% &\qquad(\zipopn{n}{\zup}\compr \arrcatch\app (\zipopn{n-1}{\enumnext}\compr \nextnoupn{n-1})\\
% &\qquad\dots\\
% &\qquad(\zipopn{1\modplus 2}{\zup}\compr \arrcatch\app (\zipopn{1\modplus 1}{\enumnext}\compr \nextnoupn{1\modplus 1})\dotsb)
% \end{align}}

\section{Discussion}\label{sec:related-work}

Our work combines approaches from the areas of (white-box) proof search
for proof assistants
and software design for functional programming languages.
One ambitious white-box proof search tool is Lean's \aesop~\cite{aesop}.
It uses an explicit AND/OR tree to implement a best-first proof search with priorities.
Its search tree can be analysed as data in Lean itself.
It offers so-called rule builders
to register tactics and theorems as search steps
and integrates several other features,
like normalisation phases and tactic script generation~\cite{aesopscripts}.
ACL2~\cite{waterfall} includes a switch to log proof trees attempted by its \waterfall as raw text.
Unlike \Zippy, both are not general frameworks for tree-based proof search
but concrete proof search implementations:
they are tied to a specific search tree datatype, are non-extensible by users,
and offer no general specification for and generation of search tree models.
As a generic framework, \Zippy does not integrate application-specific features,
like priorities or tactics as search steps,
but offers means to readily integrate them if needed,
as demonstrated in \cref{sec:exmpl-applications}.

There are various other concrete proof search implementations, including
Coq's~\cite{coq} \eauto and \sauto~\cite{sauto},
Isabelle's~\cite{lcftoholauto} \auto and \autotwo~\cite{auto2}, and
PVS's \grind~\cite{grind}.
They all differ in the set of features they integrate, search strategies they employ, etc.
But to our knowledge,
none of them is extensible nor truly white-box,
meaning that they do not provide means to readily predict their proof search
and analyse it as data.

\Zippy's design is based on abstractions known from functional programming theory.
Central to \Zippy is the concept of zippers~\cite{zipper},
which we extended to the novel concepts of linked and alternating zippers for search trees.
Our definition of zippers,
using moves $\zzip,\zunzip,\zright,\zleft,\zup,\zdown$,
is inspired by Haskell's ``Scrap Your Zippers'' framework~\cite{scrapzippers},
which implements a datatype-independent zipper in the $\typefont{option}$ Kleisli category.
The programming language community developed a rich set
of abstractions to structure functional programs.
These abstractions are prominent in Haskell,
much less common in ML-languages,
and virtually non-existent in \isabelleml.
One exception is \psl~\cite{pslmonads},
whose monads suffer from the pitfall described in \cref{sec:zippy-impl}, however.
We described a generic approach using \isabelleml antiquotations that avoids this pitfall and implemented
monads~\cite{monads}, arrows~\cite{arrows},
monad transformers~\cite{monadtransformer},
and various instances thereof
for composable computations with contexts,
and lenses~\cite{lensorig}
for composable and extensible data manipulation in \isabelleml.
While there are also formal translations from Haskell's typeclasses to slight extensions of ML~\cite{modulestypeclasses,modulestypeclasses2},
they do not handle type constructor classes.

\paragraph*{Future Work}

As discussed in \cref{sec:intro}, white-box proof search tools have many merits.
\Zippy enables users to build and extend such tools in Isabelle.
One possible use case is to create white-box alternatives to existing automation, particularly Isabelle's \auto.
Another one is porting automation from other provers, such as Lean's \aesop.
We sketched one approach to do so in \cref{sec:exmpl-applications},
including a prototype in the supplementary material.
A third use case is to create new automation for areas
requiring high levels of customisability and user extensibility.
One such area is program and theorem synthesis with relational provers~\cite{transfer,transport}.
Currently, there are three such provers of non-comparable strength in Isabelle --
\isaautoref~\cite{autoref,autorefafp},
\transfer~\cite{isarref},
\transport~\cite{transportafp}.
They are all black-box and notoriously hard for beginners.
One promising path to subsume them with a more accessible tool
is to implement a white-box variant on top of \transport's theory~\cite{transport}.
Another area is emerging soft-type developments~\cite{isaset,isagst}.
Soft-type inference algorithms share many of the challenges of mentioned relational provers
and users would hence benefit from an extensible, white-box implementation.

%%
%% Bibliography
%%
%% Please use bibtex,
% \clearpage
\bibliography{bibliography}

% \appendix

\end{document}